\documentclass[12pt]{article}
\pdfoutput=1
\usepackage{epsf,amsfonts,amssymb,epsfig,amsmath,mathtools,graphics,slashed}
\usepackage{hyperref,hep,graphicx,subfig}

\makeatletter

\newcommand{\Rmnum}[1]{\expandafter\@slowromancap\romannumeral #1@}
\makeatother

\newcommand{\nn}{\notag \\}

\begin{document}

\makeatletter
\renewcommand{\theequation}{\thesection.\arabic{equation}}
\@addtoreset{equation}{section}
\makeatother

\baselineskip 18pt

\begin{titlepage}

\vfill

\begin{flushright}
Imperial/TP/2014/JG/01\\
\end{flushright}

\vfill

\begin{center}
   \baselineskip=16pt
   {\Large\bf Novel metals and insulators\\from holography}
  \vskip 1.5cm
  \vskip 1.5cm
      Aristomenis Donos$^1$ and Jerome P. Gauntlett$^2$\\
   \vskip .6cm
    \vskip .6cm
      \begin{small}
      \textit{$^1$DAMTP, 
       University of Cambridge\\ Cambridge, CB3 0WA, U.K.}
        \end{small}\\
      \vskip .6cm
      \begin{small}
      \textit{$^2$Blackett Laboratory, 
        Imperial College\\ London, SW7 2AZ, U.K.}
        \end{small}\\*[.6cm]

\end{center}

\vfill

\begin{center}
\textbf{Abstract}
\end{center}

\begin{quote}
Using simple holographic models 
in $D=4$ spacetime dimensions we construct black hole solutions
dual to $d=3$ CFTs at finite charge density with a Q-lattice deformation. 
At zero temperature we find new
ground state solutions, associated with broken translation invariance in either one or both spatial directions,  
which exhibit insulating or metallic behaviour depending on
the parameters of the holographic theory. For low temperatures and small frequencies, the real part of the optical conductivity 
exhibits a power-law behaviour. 
We also obtain an expression for the 
the DC conductivity at finite temperature in terms of horizon data of the black hole solutions.
\end{quote}

\vfill

\end{titlepage}
\setcounter{equation}{0}


\section{Introduction}

The transition between metallic and insulating behaviour is a fascinating and enduring area of study in condensed matter physics 
(e.g. \cite{edwards,gebhard,Dobrosavljevic}). A defining property of metals and insulators is whether or not they conduct at zero temperature\footnote{After a career thinking about the topic Mott
wrote: ``I've thought a lot about `What is a metal?' and I think one can only answer the question at $T=0$. There a metal conducts, and a non-metal doesn't." \cite{edwards2}.}
and hence a continuous
metal-insulating transition constitutes a quantum phase transition (e.g. \cite{sachdev}). In many systems the 
transition arises within the context of strongly correlated electrons and it is therefore natural to look for new paradigms within the context of holography.

A simple description of holographic matter at finite charge density is captured by the electrically charged AdS-RN black brane in $D$ spacetime dimensions. At zero temperature this solution interpolates between $AdS_D$ in the UV and $AdS_2\times\mathbb{R}^{D-2}$ in the IR, and this 
locally quantum critical ground state \cite{Iqbal:2011in} has a non-zero entropy density.
The AdS-RN black hole is
translationally invariant in the $\mathbb{R}^{D-2}$ directions and, correspondingly, the real part of the optical conductivity has a delta function peak at zero frequency, for all temperatures, corresponding to the system
being an ideal conductor. This infinite DC conductivity arises from the fact that momentum is conserved and there is no mechanism for dissipating the
current.

In order to obtain more realistic metallic behaviour the translation invariance can be broken
 by incorporating a 
``holographic lattice"\footnote{Alternatively, the translation invariance can 
be broken spontaneously, as in e.g. \cite{Nakamura:2009tf,Donos:2011bh,Bergman:2011rf}. Momentum dissipation can also be achieved by using massive gravity in the bulk, as in \cite{Vegh:2013sk,Davison:2013jba,Blake:2013bqa,Davison:2013txa}, or by treating charge carriers in a probe
approximation \cite{Karch:2007pd,Faulkner:2010zz,Hartnoll:2009ns}.}. This is defined to be a deformation of the dual CFT by some type of periodic potential and is described by black holes with suitably modified asymptotic behaviour in the UV. It is possible to have constructions where the lattice deformation becomes irrelevant in the IR, in the RG sense, and 
the $AdS_2\times\mathbb{R}^{D-2}$ region of the geometry in the far IR at $T=0$ is maintained.
When this is the case the low frequency behaviour of the optical conductivity at low temperatures, with the scale set by the chemical potential, is captured by analysing perturbations about the translationally invariant $AdS_2\times\mathbb{R}^{D-2}$ geometry \cite{Hartnoll:2012rj}. One finds that
the system is in a metallic phase with the DC conductivity scaling with temperature with an exponent fixed by the least irrelevant operator in the locally quantum critical theory. Furthermore, for $T\ne 0$, the optical conductivity is expected to have a Drude-peak instead of a delta function. 
Various constructions of black holes describing holographic lattices have been made 
\cite{Horowitz:2012ky,Horowitz:2012gs,Horowitz:2013jaa,Donos:2012js,Ling:2013nxa,Chesler:2013qla,Donos:2013eha}
and Drude-peaks associated with coherent metallic behaviour are indeed 
observed\footnote{In the interesting recent solutions \cite{Andrade:2013gsa} 
(see also \cite{Bardoux:2012aw}) the translation invariance is broken
by massless scalars, without being periodic. Furthermore, the scalars are {\it marginal} deformations of the $AdS_2\times\mathbb{R}^{D-2}$ geometry.}. 
In addition,
the scaling of the DC conductivity that was predicted in \cite{Hartnoll:2012rj} was recently confirmed in \cite{Donos:2013eha} by constructing black hole solutions down to very low temperatures. 
It is worth emphasising that at $T=0$, the arguments of \cite{Hartnoll:2012rj} suggest that for these particular metallic phases, 
governed by the locally quantum critical
theory, the Drude-peak will be replaced by a delta function.
One of the results of this paper will be the construction of different metallic ground states at $T=0$,
with vanishing entropy density, where the real part of the
optical conductivity diverges
with a power law-like behaviour for small frequency or, in some models, approaches a finite value.

The potential utility of using holographic lattices for studying transitions from metallic to insulating behaviour, and also using them to find 
new insulating ground states,
was discussed in \cite{Donos:2012js}.
The key idea is to increase the strength of the holographic lattice aiming to modify the IR properties of the $T=0$ ground state. 
In \cite{Donos:2012js} 
this idea was explored
within the context of $D=5$ black holes with helical lattice deformations. For small deformations it
was shown that the system is in a metallic phase with the $T=0$ black holes interpolating between $AdS_5$ in the UV and
$AdS_2\times\mathbb{R}^3$ in the IR. For larger deformations
it was shown that there is a transition to insulating behaviour, with the corresponding $T=0$ black holes approaching new 
IR configurations with broken translation invariance.
For the insulating phase, the real part of the 
optical conductivity was shown to vanish at small frequencies, with a power law fixed by the IR geometry.
It was also shown that there is an RG unstable bifurcating solution, similar to \cite{Iizuka:2012iv}, 
which separates the
two phases.

In this paper we carry out related investigations using the framework of holographic Q-lattices,
which were
recently introduced in \cite{Donos:2013eha}.
This framework utilises a classical gravitational theory with a global symmetry in order to construct black hole solutions 
that break translation invariance and yet have a metric which is co-homogeneity one. In particular, the black hole solutions
can be obtained by solving ODEs instead of PDEs which considerably simplifies the analysis. In \cite{Donos:2013eha} a 
class of $D=4$ models with a metric, gauge-field and a neutral complex scalar field were studied.
Anisotropic black hole solutions that break translation invariance in one of the two spatial directions of the dual field theory were constructed and
shown to exhibit a
transition from metallic to insulating behaviour. 
The metallic phase was described by black holes whose $T=0$ limit 
interpolates between $AdS_4$ in the UV and $AdS_2\times\mathbb{R}^2$ in the IR. Insulating black holes were
also constructed down to very low temperatures, but the zero temperature ground states, which have zero entropy, are currently obscure.
In particular, the solutions do not seem to approach any simple power-law type
behaviour. While it would be interesting to further understand the low temperature
behaviour of the models studied in \cite{Donos:2013eha}, one might expect that simple extensions to the models 
will lead to more accessible low-temperature behaviour.

Here, in sections \ref{sec2} and \ref{sec3}, we will consider $D=4$ models with the same field content 
as those studied in \cite{Donos:2013eha} but with additional couplings. 
We will find a new class of anisotropic solutions, associated with broken translation invariance 
in one of the two spatial directions\footnote{Ground state solutions at finite charge density
which break the euclidean spatial symmetry to a helical symmetry have been found in 
\cite{Iizuka:2012iv,Donos:2012gg,Iizuka:2012pn,Donos:2013woa}.} and vanishing entropy density, 
and show that they arise as the IR limit of black hole solutions at $T=0$ and non-vanishing chemical potential $\mu$. 
Given that the $T=0$ black holes
break translation invariance one might expect that they are associated with insulating behaviour, and for certain parameters 
we find this to be the case. However, for other ranges of parameters we find that the $T=0$ ground states exhibit novel metallic behaviour. 
More precisely we will show that the real part of the conductivity 
behaves as 
$Re(\sigma)\sim \omega^y$, as $\omega\to 0$,
where $y$ is a parameter fixed by the $D=4$ theory\footnote{We note that similar power-law behaviour for the optical conductivity at low frequencies was discussed for $D=4$ holographic metals in
\cite{Donos:2012ra} and has also been seen for $D=5$ helical metals and insulators \cite{to app}.}. We will also show that the DC conductivity scales like $\sigma_{DC}\sim T^{y}$
for $T<<\mu$ and $T<<k$. Thus the ground states in models with $y>0$ correspond to
insulating behaviour, while in models with $y<0$ they correspond to metallic behaviour. 
It should be emphasised that these incoherent holographic metals, with ground states which break translation invariance,
are quite distinct from the coherent holographic metals which have ground states built from irrelevant deformations of the
translationally invariant $AdS_2\times\mathbb{R}^2$ solution. In particular, our new metallic ground states are not amenable to being studied via the 
memory matrix formalism described in \cite{Hartnoll:2012rj}.
There are also models with $y=0$, leading to a constant DC conductivity at $T=0$.
This latter type of behaviour was also seen in \cite{Davison:2013jba,Andrade:2013gsa} but, in contrast to what we find here,
there it was in the context of non-vanishing entropy density and also with a temperature independent DC conductivity.

In section \ref{sec4} of the paper, we consider a slightly more general class of models which
includes a real scalar field with two real ``axion" fields. We show that the model admits new solutions associated with ground states
that break translation invariance in {\it all} of the spatial directions of the dual field theory. 
We show that the ground states can again exhibit novel insulating and metallic
behaviour. We now find examples where the optical conductivity and the DC conductivity exhibit scaling
behaviours with {\it different} exponents.
We also find new $AdS_2\times\mathbb{R}^2$ solutions supported by linear axion fields with broken translation invariance.
For a closely related set of models, with massless
axion fields, we show in an appendix that they give rise to a novel generalisation of ``$\eta$-geometries" \cite{Hartnoll:2012wm,Donos:2012yi}. 
Recall that the $\eta$-geometries are conformal to $AdS_2\times\mathbb{R}^2$
and have been emphasised as a framework for holographically studying semi-local quantum criticality in \cite{Hartnoll:2012wm}.

In section \ref{sec5} we present our calculation of  
the DC conductivity, $\sigma_{DC}$, for a general class of finite temperature black hole solutions using a new approach.
Instead of taking a zero-frequency limit of the optical conductivity, as is usually done, we switch on a time 
independent electric field from the start and then determine the response. By carefully analysing regularity conditions we 
obtain an elegant expression for the DC conductivity in terms of horizon data. This complements similar results for specific models 
in a translationally invariant context \cite{Iqbal:2008by} and in the context of momentum dissipation
\cite{Blake:2013bqa,Blake:2013owa,Andrade:2013gsa}.

\section{Anisotropic ground states which break translation invariance in one spatial direction}
\label{sec2}

Consider a $D=4$ model of gravity coupled to a gauge-field $A$ and a neutral complex scalar field $z$ with action given by
\begin{align}\label{act1}
S=\int d^4 x\sqrt{-g}\left[R-V(|z|)-\frac{1}{4}\tau(|z|)F^2-G(|z|)|\partial z|^2\right]\,,
\end{align}
where $F=dA$ and $V,\tau$ and $G$ are functions of the modulus of the complex scalar. The
model has an abelian gauge symmetry in which $A\to A-d\Lambda$, 
as well an abelian global symmetry associated with constant phase
rotations of $z$. 

We can choose $V,\tau$ and $G$ so that the model admits
an $AdS_4$ vacuum which will be dual to a $d=3$ CFT.
We are interested in studying the behaviour of this CFT when held at finite charge density
with respect to the global symmetry associated with the bulk gauge-field, and with a Q-lattice deformation. Recall
from \cite{Donos:2013eha} that the Q-lattice utilises the bulk global symmetry to break translation invariance. In the particular setting
of the model defined by \eqref{act1} the Q-lattice introduces a phase in the 
asymptotic behaviour of the complex field $z$ that depends {\it periodically} on one of the spatial directions, and hence
breaks translation invariance explicitly.

We want to identify possible zero temperature black holes of this set-up,
aiming to find ground states with novel IR behaviour.
In fact we have constructed finite temperature Q-lattice black holes and cooled them down to low-temperatures finding that they
can indeed approach such configurations. We will describe these black hole solutions in section \ref{sec3}, but we will first
focus on describing the ``fixed point solutions" solutions that capture the IR behaviour of the ground states. 
As we will see these solutions involve simple power-law like behaviour for the fields and they capture key low-energy features
of the conductivity of the full $AdS_4$ black holes. It is worth noting in advance that they are singular and moreover have running scalar fields.

We will study models in which the scalar manifold appearing in \eqref{act1}, with coordinate $z$ and line element
$G(|z|)|dz|^2$, is non-compact and asymptotically, locally approaches hyperbolic space at the boundary, with real coordinates
$\phi,\chi$ and line element $d\phi^2+e^{2\phi}d\chi^2$. Furthermore, in the IR region of the $T=0$ black hole solutions we will see that
the scalar field is driven to this
boundary $\phi\to\infty$. In addition, we find, that the ground state solutions are determined by
the behaviour of the functions $V,\tau$ depending on exponentials of $\phi$.

Assembling these ingredients, we therefore focus on finding fixed point solutions of the following class\footnote{Related models have been studied in other holographic contexts e.g. 
\cite{Goldstein:2009cv,Cadoni:2009xm,Charmousis:2010zz,Mateos:2011tv,Iizuka:2012wt}.} 
of models:
\begin{align}\label{act2}
S=\int d^4 x\sqrt{-g}\left[R-\frac{c}{2}\left[(\partial\phi)^2+e^{2\phi}(\partial\chi)^2\right]+n_1e^{\alpha\phi}-n_2\frac{e^{\gamma\phi}}{4}F^2\right]\,,
\end{align}
where $c>0,\alpha$ and $\gamma$ are constants. For simplicity, 
we will shortly focus on the one-parameter family of models labelled by $\gamma$ by setting $c=3,\alpha=1$.
We have also included two additional constants, $n_1$ and $n_2>0$, which will
play a minor role, as one can see by rescaling the metric and/or the gauge-field $A$,
but are added for later convenience.
Observe that the global symmetry of the bulk theory corresponds to constant shifts of the ``axion" field $\chi$.
The ansatz that we consider is given by
\begin{align}\label{ansatzbh}
ds^2&=-Udt^2+U^{-1}dr^2+e^{2V_1}dx_1^2+e^{2V_2}dx_2^2\,,\nn
A&=adt\,,\nn
\chi&= kx_1\,,
\end{align}
with $U,V_1,V_2,a$ and $\phi$ functions of $r$. 
In the black hole setting discussed in section \ref{sec3}, $\chi$ will be a periodic field and hence we
see that this ansatz is associated with breaking of translation invariance in the $x_1$ direction, whilst preserving it in
the $x_2$ direction. Generically the metric is cohomogeneity one and anisotropic in the $x_1,x_2$ directions.
We also note that this lattice breaks parity.
The equations of motion lead to a system of differential equations which are first order in
$U$ and second order in $V_1,V_2,a$ and $\phi$. Notice that the ansatz \eqref{ansatzbh} is left invariant under the separate
scaling symmetries given by
\begin{align}
&x_1\to \kappa x_1,\quad e^{V_1}\to \kappa^{-1} e^{V_1},\quad k\to \kappa^{-1}k;\label{scal1}\\
&x_2\to \kappa x_2,\quad e^{V_2}\to \kappa^{-1} e^{V_2};\label{scal2}\\
&t\to \kappa t,\quad r\to \kappa^{-1}r,\quad U\to \kappa^{-2} U,\quad a\to \kappa^{-1} a\,;\label{scal3}
\end{align}
for constant $\kappa$.

We now look for new fixed point solutions of the form
\begin{align}\label{gdstate}
U=L^{-2}r^{u_1},\quad e^{V_1}=e^{v_{10}}r^{v_{11}},\quad e^{V_2}=e^{v_{20}}r^{v_{22}},\quad a=a_0 r^{a_1},\quad e^{\phi}=e^{\phi_0}r^{\phi_{1}}\,.
\end{align}
After substituting into the equations of motion we obtain algebraic equations for ten constants.
These imply that the radial exponents are given by
\begin{align}\label{exps}
u_1&=\frac{2 \left(8+2 {c}-4 \alpha +4 \gamma -\alpha  \gamma +\gamma ^2\right)}{8+2 {c}-4 \alpha +\alpha ^2+4 \gamma +\gamma ^2}\,,\nn
v_{11}&=\frac{(-2+\alpha ) (\alpha +\gamma )}{8+2 {c}-4 \alpha +\alpha ^2+4 \gamma +\gamma ^2}\,,\nn
v_{22}&=\frac{(2+\gamma ) (\alpha +\gamma )}{8+2 {c}-4 \alpha +\alpha ^2+4 \gamma +\gamma ^2}\,,\nn
a_{1}&=\frac{2 \left(4+{c}-2 \alpha +2 \gamma +\gamma ^2\right)}{8+2 {c}-4 \alpha +\alpha ^2+4 \gamma +\gamma ^2}\,,\nn
\phi_1&=-\frac{2 (\alpha +\gamma )}{8+2 {c}-4 \alpha +\alpha ^2+4 \gamma +\gamma ^2}\,,
\end{align}
and this leads to an algebraic equation for the coefficients. While solutions exist for various values of $\alpha$, we just record the solution
for $\alpha=1$ which is of most interest:
\begin{align}\label{scales}
L^2&=\frac{4 {c} k^2 (2+{c}+2\gamma  +\gamma^2) (3+{c}+3\gamma  +\gamma^2)^2}{{n_1}^2 e^{2 {v_{10}}} (1+\gamma ) (3+\gamma ) (5+2 {c}+4\gamma  +\gamma^2)^2}\,,\nn
a_0&=\pm\frac{\sqrt{n_1(1+{c}) }  (5+2 {c}+4\gamma +\gamma^2)e^{\frac{1-\gamma}{2} {\phi_0}}}{\sqrt{2 n_2} (2+{c}+2\gamma  +\gamma^2) \sqrt{3+{c}+3\gamma  +\gamma^2}}\,,\nn
e^{\phi_0}&=\frac{{n_1} e^{2 {v_{10}}} (1+\gamma ) (3+\gamma )}{{c} k^2 (3+{c}+3\gamma  +\gamma^2)}\,.
\end{align}
Notice that the constants $n_1,n_2$ only appear in \eqref{scales} and do not appear in the exponents \eqref{exps}. 
We also see that $n_1>0$. Similarly $k$ and $e^{v_{10}}$
appear in the combination $ke^{-v_{10}}$ and again only in \eqref{scales}.

To simplify the presentation we will now restrict to a one-parameter family
of models, labelled by $\gamma$,  by setting
\begin{align}\label{vale}
c=3,\qquad \alpha=1,\qquad \gamma>-1\,.
\end{align}
We then have $L^2,e^{\phi_0}>0$, 
as required for the fixed point solutions to exist, $u_{1},v_{22},a_{1}>0$ and $v_{11}, \phi_{1}<0$.
In particular, we see that as $r\to 0$ we have $e^\phi\to\infty$ as we had assumed in the discussion
preceding \eqref{act2}.

\subsection{Mode analysis}\label{sec:mode_an}

We next consider, within our ansatz \eqref{ansatzbh}, 
the static perturbations about the new fixed point solutions.
We consider
\begin{align}
U&=L^{-2}r^{u_1}(1+c_1 r^\delta),\quad e^{V_1}=e^{v_{10}}r^{v_{11}}(1+c_2r^\delta),
\quad e^{V_2}=e^{v_{20}}r^{v_{22}}(1+c_3r^\delta),\nn
 a&=a_0r^{a_1}(1+c_4 r^\delta),\quad e^{\phi}=e^{\phi_0}r^{\phi_{1}}(1+c_5 r^\delta)\,.
\end{align}
for small $c_i$. 
After substituting into the equations of motion and keeping terms linear in the $c_i$, we find
a single solution with $\delta=-1$, which just corresponds to shifting $r\to r-r_+$, for constant $r_+$, in the solution \eqref{gdstate}. In addition we find four pairs of solutions of the form
$\delta_\pm$, each with $\delta_+\ge \delta_-$ and $\delta_++\delta_-=1-u_1-v_{11}-v_{22}$.
Two of these pairs are marginal modes with $\delta_+=0$ and arise
from the scaling symmetries \eqref{scal2}, \eqref{scal3}. Specifically, \eqref{scal3} implies that if $U(r),V_i(r),a(r),\phi(r)$ solve the equations of motion at fixed $k$ then so will $c^{-2}U(cr),V_i(cr),c^{-1}a(cr),\phi(cr)$.
The remaining two pairs have
\begin{align}
\delta^{(\pm)}_+&=-\frac{5+2 \gamma+ \gamma ^2}{\left(11+4 \gamma +\gamma ^2\right)}\nn
&+\frac{
 \left(5+2 \gamma +\gamma ^2\right)^{1/2} \left(87+46 \gamma +19 \gamma ^2
\pm 8 \sqrt{27-9 \gamma -2 \gamma ^2-\gamma ^3+\gamma ^4}\right)^{1/2}}{\sqrt{3} \left(11+4 \gamma +\gamma ^2\right)}
\end{align}
One can check that for $\gamma>-1$ we have $\delta^{(\pm)}_+>0$ and hence correspond to
RG irrelevant modes. 
Combining these with the two $\delta_+=0$ modes as well  
as the $\delta=-1$ solution we can develop an IR expansion with five parameters. In section \ref{sec3} we will discuss an embedding 
of these solutions into a specific class of theories which have an $AdS_4$ vacuum. The five IR parameters
are a sufficient number to be able, generically, to construct 
$T=0$ domain wall solutions interpolating between the new fixed point solutions in
the IR and 
a Q-lattice deformed $AdS_4$ in the UV.

It is helpful to also note that there is a static perturbation corresponding to creating a small black hole.
Specifically, associated with the two $\delta_{+}=0$ solutions there is a solution with
$\delta_-=1-u_{1}-v_{11}-v_{22}\equiv \delta_{T}$ and 
$c_{1}=c_{4}\equiv-\epsilon$, $c_{2}=c_{3}=c_{5}=0$. This leads to 
a regular Killing horizon at $r=\epsilon^{-1/\delta_{T}}\equiv r_+$ 
with temperature $T\propto r_{+}^{u_{1}-1}$ and entropy density $s\propto r_{+}^{v_{11}+v_{22}}$. 
We are interested in realising the fixed point solutions as the near horizon, zero temperature limit 
of black holes which approach these small black holes in this limit. Now,
for the fixed point solutions we are considering, with \eqref{vale} and $\gamma>-1$, we can check that $u_{1}>1$ and $v_{11}+v_{22}>0$. The condition $u_{1}>1$ ensures that the surface gravity of the small black hole solutions
goes to zero as $r_{+}\rightarrow 0$. The condition $v_{11}+v_{22}>0$ implies that the entropy density goes to
zero, and the two conditions together imply that 
the entropy density is an increasing function of the temperature, giving a positive specific heat as needed
for thermodynamic stability.

\subsection{Conductivity}\label{sec:cond}

In order to analyse the optical conductivity we consider\footnote{Note that the optical conductivity in the $x_2$ direction will have a delta function at finite temperature due to the translation invariance in this direction.} the following consistent linear perturbation about the fixed point solutions \eqref{gdstate}:
\begin{align}
g_{tx_1}&=\delta h_{tx_1}(t,r)\,,\nn
A_{x_1}&={\delta a_{x_1}}(t,r)\,,\nn
\chi&=kx_1+\delta\chi(t,r)\,,
 \end{align}
 where $\delta h_{tx_1},\delta a_{x_1},\delta\chi$ are real.
 We then allow for time dependence by setting
 \begin{align}
\delta h_{tx_1}&=e^{-i\omega t}\delta h_{tx_1}(r),\nn
 {\delta a_{x_1}}&=e^{-i\omega t}{\delta a_{x_1}}(r)\,,\nn
 \delta\chi&=ie^{-i\omega t}\delta\chi(r)\,.
 \end{align}
Substituting into the equations of motion leads to
 \begin{align}
&\delta a_{x_1}''{}+
\left(\gamma  \phi '{}+\frac{ U'{}}{U{}}-{V_1}'{}+ {V_2}'{}\right)\delta a_{x_1}' 
+\left(\frac{\omega ^2{}}{U{}^2} -\frac{{n_2} {} a'{}^2 e^{\gamma  \phi {}}}{U{}}\right){\delta a_{x_1}}
+\frac{ {c} k e^{2 \phi {}} a'{} {}}{\omega }\delta \chi '=0\,,\label{one}\\
&\delta \chi ''{}
+\left(2 \phi '{}+\frac{ U'{}}{U{}}+ {V_1}'{}+ {V_2}'\right)\delta\chi'+\frac{\omega ^2  {}}{U{}^2}\delta \chi
-\frac{ k \omega  {} e^{-2 {V_1}{}}}{U{}^2}{\delta h_{tx_1}}=0\,,\label{two}\\
&{n_2} {} a'{} e^{\gamma  \phi {}}{\delta a_{x_1}}-\frac{ {c} k U{} e^{2 \phi {}} }{\omega }\delta \chi '{}+{\delta h'_{tx_1}}-2 {} {V_1}'{}{\delta h_{tx_1}}=0\,.\label{three}
\end{align}
 
We can solve \eqref{two} for ${\delta h_{tx_1}}$ and substitute into \eqref{three}.
Next, after substituting in the ground state solution and defining
\begin{align}
\Phi_1&=r^{\frac{\gamma ^2+8 \gamma +23}{2 \left(\gamma ^2+4 \gamma +11\right)}}\delta a_{x_1}\,,\nn
\Phi_2&=\omega^{-1}r^{\frac{7 \gamma ^2+16 \gamma +57}{2 \left(\gamma ^2+4 \gamma +11\right)}}\delta \chi '\,,
\end{align}
we obtain differential equations that can be written in the form
\begin{align}\label{beseq}
\left(\partial_r^2+x\omega^2 r^{-\frac{4 \left(\gamma ^2+3 \gamma +10\right)}{\gamma ^2+4 \gamma +11}}+r^{-2}M\right)
\begin{pmatrix}
\Phi_1\\\Phi_2
\end{pmatrix}
=0\,.
\end{align}
Here $x$ is a constant given by
\begin{align}
x=\frac{144 \left(5+2\gamma+\gamma ^2\right)^2 \left(6+3\gamma+\gamma ^2\right)^4 k^4 e^{-4 {V_{1}} }}{{n_1}^4 (1+\gamma )^2 (3+\gamma)^2 \left(11+4\gamma+\gamma ^2\right)^4}\,,
\end{align}
and $M$ is a two by two matrix with eigenvalues given by
\begin{align}
M_1&=-\frac{(21+8\gamma+3\gamma^2)  (43+16\gamma+5\gamma^2)}{4(11+4\gamma+\gamma^2)^2},\nn
M_2&=\frac{(\gamma^2 -1) (23+8\gamma+\gamma^2)}{4(11+4\gamma+\gamma^2)^2}\,.
\end{align}

Observe that \eqref{beseq} only depends on the frequency via $\omega^2$ and hence the Greens
functions and the conductivity will be symmetric functions of $\omega$. We will continue assuming that
$\omega\ge 0$.
The general solution of the differential equation \eqref{beseq},
with either eigenvalue $M_i$, is expressed in terms of Hankel functions:
\begin{align}
\Psi_{i}=r^{1/2}\,c_{i}^{(1)}\,H^{(1)}_{\frac{\left(11+4\gamma+\gamma^{2} \right)\,\sqrt{1-4 M_{i}}}{2\left(9+2\gamma+\gamma^{2}\right)}}\left(c_{0} \omega\,r^{-\frac{9+2\gamma+\gamma^{2}}{11+4\gamma+\gamma^{2}}}  \right)+r^{1/2}\,c_{i}^{(2)}\,H^{(2)}_{\frac{\left(11+4\gamma+\gamma^{2} \right)\,\sqrt{1-4 M_{i}}}{2\left(9+2\gamma+\gamma^{2}\right)}}\left(c_{0} \omega\,r^{-\frac{9+2\gamma+\gamma^{2}}{11+4\gamma+\gamma^{2}}}  \right)
\end{align}
with $c_0=\frac{(11+4\gamma+\gamma^2)}{(9+2\gamma+\gamma^2)}\sqrt{x}$.

To obtain the retarded Greens functions ${\cal G}_{\Psi_{i}\Psi_{i}}$ for the fixed point solutions, 
we next need to impose in falling boundary conditions as $r\to 0$.
For 
\eqref{ansatzbh} the tortoise coordinate is $r_*=r^{1-u_1}/(L(1-u_1)$ and we want solutions
which behave as $e^{-i\omega \left(t+r_*\right)}$. Observing that 
$r_*\sim r^{-\frac{\gamma ^2+2 \gamma +9}{\gamma ^2+4 \gamma +11}}$. We thus conclude that, for $\omega\ge 0$, we need to choose $c_{i}^{(1)}=0$. Expanding the functions at $r\rightarrow\infty$ we then have
\begin{align}
\Psi_{i}=d_{i}\left(r^{\frac{1}{2}(1+\sqrt{1-4M_i})}+\ldots+{\cal G}_{\Psi_{i}\Psi_{i}}\,r^{\frac{1}{2}(1-\sqrt{1-4M_i})}+\ldots\right)\,,
\end{align}
for some constants $d_i$, and we obtain the scaling
\begin{align}\label{ircf}
{\cal G}_{\Psi_{1}\Psi_{1}}\propto\,\omega^{\frac{4\,\left( 8+3\gamma+\gamma^{2}\right)}{9+2\gamma+\gamma^{2}}},\qquad 
{\cal G}_{\Psi_{2}\Psi_{2}}\propto\,\omega^{\frac{4\,\left( 3+\gamma\right)}{9+2\gamma+\gamma^{2}}}\,.
\end{align}
This scaling can also be obtained by solving the
$\omega=0$ equations \eqref{beseq} and then using the fact that
$\omega$ only appears in \eqref{beseq} in the combination $\omega/r^\frac{9+2\gamma+\gamma^2}{11+4\gamma+4\gamma^2}$.

In the next section we will construct black hole solutions which approach $AdS_4$ in the UV and approach the new 
fixed point solutions in the far IR at $T=0$. As explained in \cite{Donos:2012ra}, generalising arguments in
\cite{Faulkner:2009wj}, a matching procedure can be used to show that the low-frequency behaviour of 
the imaginary part of the UV two point functions at low frequencies is determined by the imaginary part of the IR correlation functions. From \eqref{ircf} we see that the dominant contribution for small $\omega$ will be
given by the solution $\Psi_2$ and hence the contribution to the real part of the optical conductivity for low frequencies will be given by
\begin{align}\label{conscal}
\mathrm{Re}\sigma\propto \frac{1}{\omega}\mathrm{Im}{\cal G}_{\Psi_{2}\Psi_{2}}\propto\omega^{\frac{(1+\gamma)(3-\gamma)}{9+2\gamma+\gamma^2}}\,.
\end{align}

Taking the $\omega\to 0$ limit of\eqref{conscal} we conclude that
for $-1<\gamma<3$ the conductivity is going to zero and we have insulating behaviour. 
For $\gamma>3$ the DC conductivity has a power-law divergence and we have metallic behaviour.
For $\gamma=3$ we have an intermediate behaviour where the DC conductivity approaches a constant.

Actually, some care is required here since in addition to the power-law, $\sigma(\omega)$ might also have
delta function contributions of the form $\delta(\omega)$. 
While such delta functions can be inferred, via Kramers-Kr\"onig
relations, from the existence of poles in the imaginary part of the optical conductivity, this is not a 
straightforward calculation
in the present setting. However, we can obtain some insight by calculating the 
DC conductivity at finite temperature. In section \ref{sec5} we show that as $T\to 0$ we have the scaling behaviour
 \begin{align}
 \sigma_{DC}\sim T^{\frac{(1+\gamma)(3-\gamma)}{9+2\gamma+\gamma^2}}\,,
 \end{align}
exactly in accord with \eqref{conscal}.
Note that $\sigma_{DC}$ is finite for $-1<\gamma\le 3$ implying that no delta functions are present, at least for these values.

\subsection{Scaling and finite temperature}

In section \ref{sec3} we will construct black holes whose zero temperature limit approaches the 
fixed point solutions \eqref{gdstate} in the IR.
Using an argument in \cite{Huijse:2011ef} we can deduce the temperature scaling of the entropy entropy density of these black holes.
We first observe that under the scaling 
\begin{align}
t\to \zeta^{-\frac{9+2\gamma+\gamma^2}{(1+\gamma)}}t,\quad
x_1\to \zeta^{2}x_1\quad
x_2\to\zeta^{-(1+\gamma)}x_2,\quad
r\to\zeta^\frac{11+4\gamma+\gamma^2}{(1+\gamma)}r\,,
\end{align}
the metric scales via $ds\to \zeta ds$. 
Note that we also have $e^{\phi}\to\zeta^{-2}e^\phi$, and the one-form $A$ scales as $A\to \zeta^{1+\gamma}A$. Furthermore, without scaling $k$, the action given in \eqref{act2} scales as $S\to\zeta^2S$.
This is like an anisotropic version of the solutions dual to hyperscaling violation in \cite{Charmousis:2010zz,Ogawa:2011bz,Huijse:2011ef}.
Now, for very low temperature black holes the entropy density, $s$, will scale in the same way as  $r^{v_1+v_2}$.
Since the temperature $T$ will scale in the same way as $t^{-1}$ we deduce that as $T\to 0$ we should have the scaling
\begin{align}\label{enscal}
s\sim T^\frac{(1+\gamma)^2}{9+2\gamma +\gamma^2}\,,
\end{align}
with, in particular, $s\to 0$ as $T\to 0$.
By a similar argument we can also conclude that the scalar field $\phi$ and the sure of the field strength, $F^2$, scale as
\begin{align}\label{dilscal}
e^\phi\sim T^{-\frac{2(1+\gamma)}{9+2\gamma +\gamma^2}},\qquad
F^2\sim T^{\frac{2(\gamma^2-1)}{9+2\gamma +\gamma^2}}\,.
\end{align}

An alternative and more direct way to obtain \eqref{enscal} and \eqref{dilscal} is to use
the static mode presented towards the end of section \ref{sec:mode_an} to construct a small black hole and then deduce the scaling from there.

\section{Black holes in a UV completion}\label{sec3}
We would like to realise the fixed point solutions that we discussed in the last section as IR ground states of
black hole solutions with $AdS_4$ asymptotics in the UV
in the zero temperature limit. A priori this is not guaranteed. Having a sufficiently large number of marginal and  
irrelevant deformations in the ground states combined with a sufficient 
number of marginal and relevant modes in the UV, provides very
good evidence that $T=0$ domain wall solutions will exist,
and that they can be heated up to finite temperature. However, it is not clear that the black holes can be heated up to arbitrarily large temperatures
as this will depend on the details of the gravity theory. Indeed there are examples of $AdS_2\times\mathbb{R}^2$ ground states, for example,
which can only heated up to some maximal temperature (e.g. \cite{Donos:2012yu}). 

For the UV completion we will keep the same field content of metric, gauge field and scalar fields $\phi, \chi$. This still leaves 
a great deal of freedom in choosing a particular model to analyse. We have found that the following model, which 
arose from adapting some numerics for a model in another context, has several appealing features and is the
one we shall consider:
\begin{align}
S=\int d^4x\sqrt{-g}\left[R-\frac{1}{4}\cosh^{\gamma/3}(3\phi)F^2+6\cosh\phi-\frac{3}{2}(\partial\phi)^2-{6}\sinh^2\phi(\partial\chi)^2\right].
\end{align}
Observe that the for the scalar manifold to be smooth at $\phi=0$, the axion field $\chi$ should be a periodic field, with period $\pi$.

This model admits an $AdS_4$ vacuum with unit radius. The field $\phi$ is dual to an operator with dimension $\Delta=2$ and
the massless field $\chi$ is dual to a marginal operator with dimension $\Delta=3$. We also note that as $\phi\to\infty$ then we obtain \eqref{act2}
with $c=3,\alpha=1$ and in addition $n_1=3$, $n_2=2^{-\gamma/3}$. 

This model also admits the standard electrically charged AdS-RN black hole solution, with $\phi=\chi=0$.
Provided that this solution
is dynamically stable (and that there are no first-order transitions to other possible branches of black hole solutions), it will  
describe the dual translationally invariant CFT at finite $T,\mu$.
At $T= 0$ it approaches 
the following $AdS_2\times\mathbb{R}^2$ solution in the far IR:
\begin{align}\label{ads2}
ds^2&=\frac{1}{6}ds^2(AdS_2)+dx_1^2+dx_2^2\,,\nn
F&=\frac{1}{\sqrt{3}}Vol(AdS_2)\,,
\end{align}
where $ds^2(AdS_2)$ denotes the standard unit radius metric on $AdS_2$. 

The Q-lattice black holes, to be described below, lie within the ansatz given in \eqref{ansatzbh}. 
It will be helpful to analyse the following linearised perturbations\footnote{Note that since $\phi=0$ in
\eqref{ads2} we can add $\chi=k x_1$ for finite $k$.} 
about the $AdS_2\times\mathbb{R}^2$
solution, that lie within this ansatz:
\begin{align}
U&=6r^2(1+u_1 r^\delta),\quad V_1=v_{10}(1+v_{11}r^\delta),\quad V_2=v_{20}(1+v_{21}r^\delta),\nn
 a&=2\sqrt{3}r(1+a_1 r^\delta),\quad\phi=\phi_1 r^\delta,\quad \chi=k x_1\,.
\end{align}
The corresponding perturbations are associated with operators with scaling dimension $\Delta=-\delta$ or $\Delta=\delta+1$
in the locally quantum critical IR theory that is captured by the $AdS_2\times\mathbb{R}^2$ solution. 
We find after substituting into the equations of motion a single solution with $\delta=-1$ corresponding to shifting $r\to r-r_+$ in \eqref{ads2},
and an additional four pairs of solutions, each satisfying $\delta_+>\delta_-$ and $\delta_++\delta_-=-1$. There are two marginal modes with $\delta_+=0$, corresponding
to scaling $x_1,x_2$ in \eqref{ads2} and another mode with $\delta_+=1$. Finally there is a mode depending on $k$, and just
involving the scalar field, with $\delta_+=\delta_\phi$ where
\begin{align}\label{delphi}
\delta_\phi=-\frac{1}{2}+\frac{1}{6}  \sqrt{24 e^{-2v_{10}}k^2 -3 (12 \gamma +1) }\,.
\end{align}

When $\delta_\phi\ge 0$ we can combine it with the modes with $\delta=0,0,1$ and the solution with $\delta=-1$ corresponding to shifts of $r$, 
to develop an IR
expansion to the equations of motion that depends on five parameters. This is a sufficient number to be able, generically,
to construct a domain wall solution interpolating between $AdS_2\times\mathbb{R}^2$s in the IR and 
a Q-lattice deformed $AdS_4$ in the UV, to be discussed in the next sub-section.

Observe that \eqref{delphi} is minimised for $k=0$ and also that $\delta_\phi\ge 0$ provided that $2 e^{-2v_{10}}k^2\ge 1+3\gamma$.
It is worth emphasising, as in \cite{Donos:2012js,Donos:2013eha}, that the parameter $v_{10}$ is fixed by the full domain wall solution and depends on the UV data.
We thus see that for $-1<\gamma\le -1/3$ we always have $\delta_\phi> 0$ at $k\ne 0$
and hence the
$AdS_2\times\mathbb{R}^2$ solution is always RG stable i.e. it has no relevant or marginal
deformations. 
When $-1/3<\gamma\le -1/12$ we see that the $AdS_2\times\mathbb{R}^2$ solution is RG stable when 
$2 e^{-2v_{10}}k^2> 1+3\gamma$ but is RG unstable or marginal
when $2 e^{-2v_{10}}k^2\le 1+3\gamma$. Finally, when
$-1/12<\gamma$ we see that at $k=0$, $\delta_\phi$ is complex and hence the corresponding mode violates the BF bound leading to
a dynamical instability of the $AdS_2\times\mathbb{R}^2$ solution and hence also the AdS-RN solution.  
We will return to these points in the next subsection.

\subsection{Q-lattice black holes}
The ansatz for the Q-lattice black holes that we shall consider is again given by \eqref{ansatzbh}. It is worth emphasising that
since the field $\chi$ is periodic, the black hole solutions are periodic, with $x_1=x_1+\pi/k$, and hence define a holographic lattice.
The equations of motion lead to a system of differential equations that are first order 
in $U$ and second order in $V_1,V_2,a$ and $\phi$. The black hole event horizon is taken to be at $r=r_+$ and we can develop
the following expansion near $r=r_+$:
\begin{align}
U&=4\pi T(r-r_+)+\dots,\nn
V_1&=V_{1+}+z_1V_{22}(r-r_+)\dots, \nn
V_2&=V_{2+}+V_{22}(r-r_+)\dots, \nn
 a&=a_+(r-r_+)+z_2V_{22}(r-r_+)^2\dots,\nn
\phi&=\phi_++z_3V_{22}(r-r_+)\dots,
\end{align}
where $T$ is the temperature of the black hole given by
\begin{align}
T=(4\pi)^{-1}\frac{12 {\cosh}(\phi_+)-a_+^2 {\cosh}^{\gamma /3}(3\phi_+)}{4 V_{22}}\,,
\end{align}
and the constants $z_1,z_2$ and $z_3$ can all be expressed in terms of $V_{1+},V_{2+},a_+,\phi_+$ and $k$.
Thus, for given $k$, this expansion depends on six parameters $r_+,V_{1+},V_{2+},V_{22},a_+$ and $\phi_+$.

At the $AdS_4$ boundary in the UV we can develop the following expansion as $r\to \infty$,
\begin{align}\label{uvexp}
U&=r^2-\frac{3\lambda^2}{4}-\frac{M}{r}+\dots,\nn
 V_1&=\log r-\frac{3\lambda^2}{8r^2}+\frac{V_v}{r^3}+\dots,\nn
 V_2&=\log r-\frac{3\lambda^2}{8r^2}+\frac{-V_v-\phi_c}{r^3}+\dots,\nn
a&=\mu+\frac{q}{r}+\frac{q\lambda^2(1-2\gamma)}{4r^3}+\dots,\nn
\phi&=\frac{\lambda}{r}+\frac{\phi_c}{r^2 }+\frac{\lambda(24 k^2+7\lambda^2)}{12 r^3}+\dots\,.
\end{align}
This gives a UV expansion that depends on seven parameters $M, V_v, \mu,q,\lambda,\phi_c$ and $k$. Note, in particular,  
that $\mu$ gives the chemical
potential and $\lambda$ parametrises the strength of the Q-lattice deformation. 

A parameter count, spelled out in \cite{Donos:2013eha}, shows that we expect a three-family parameter of black hole solutions,  
which we can take to be the three dimensionless quantities: $T/\mu$, $\lambda/\mu$ and $k/\mu$. 
A detailed investigation of the full phase diagram
as these three parameters are varied is left for future work.
Our principle objective here is to construct specific examples of new black hole solutions whose $T=0$ limit
maps onto the new fixed point solutions that we constructed in the last section, in the far IR. 

Indeed we have constructed such black holes, for certain values of $\gamma$, and cooled them down to very low temperatures, $T<<\mu$, in order to
extract the $T=0$ ground states. One way to verify that the functions 
approach the power law behaviour \eqref{gdstate} close to the horizon is to plot, for example, 
$\left( r-r_{+}\right)\,\phi^{\prime}\left(r\right)$ as a function of $r$ and identify a region where this becomes a constant equal to $\phi_{1}$ in agreement with \eqref{gdstate}.
We have also checked that as $T\to 0$ the entropy of the black holes scales with temperature exactly as in \eqref{enscal}; see figure \ref{figone}. Similarly, we checked that $\phi_+$, the value of the scalar field at the black hole horizon, 
diverges as the temperature is lowered exactly as given in \eqref{dilscal} and that the horizon value of $F^2$ as a function of temperature also agrees with \eqref{dilscal}.
Finally, we have numerically calculated the optical conductivity for some of 
the full black hole solutions 
and we find that as $T\to 0$ and $\omega\to 0$, i.e $\omega<<\mu,\lambda$ but $\omega>>T$,
it approaches the behaviour given in \eqref{conscal}. Moreover, for some 
examples with insulating ground states, when $-1<\gamma<3$, we checked that
the DC conductivity is as an increasing function of temperature, for $T<<\mu$, while for the metallic ground states, when $\gamma>3$, it
is a decreasing function, as expected.

\subsubsection{$-1<\gamma\le -1/3$}
For these models the AdS-RN black hole is dynamically stable and the $AdS_2\times\mathbb{R}^2$ solution
in the IR at $T=0$ is RG stable for all $k\ne 0$ (i.e. $\delta_\phi>0$ in \eqref{delphi}). For these models 
it is not clear what to expect {\it a priori}.
We have found the following results. For $\gamma=-2/3$,
$k/\mu=\sqrt{2}/20$ and $\lambda/\mu=1/10$ we find that the Q-lattice deformation gives rise to black hole
solutions at $T=0$ which still approach the $AdS_2\times\mathbb{R}^2$ solution \eqref{ads2} in the IR. 
The fact that the ground states have non-vanishing entropy is
illustrated in figure \ref{figone}(a). These black holes will be
``conventional" holographic metals with the optical conductivity having a Drude-peak and
with DC resistivity scaling as $\rho_{DC}\sim (T/\mu)^{2\delta_\phi}$ as $T\to 0$ \cite{Hartnoll:2012rj}.

With $k/\mu$ fixed, increasing the strength of the lattice deformation to 
$\lambda/\mu=3/2$, we find black holes whose $T=0$ limit approach the new fixed point solutions in the IR, as illustrated
in figure \ref{figone}(a).  Recall from \eqref{conscal} that for $-1<\gamma\le -1/3$
these ground states are insulating and we have thus realised a metal-insulator transition.
Notice that for this case we have from \eqref{enscal} that $s\sim T^{0.014}$. 
The two curves in figure 
\ref{figone}(a) show that we have cooled the black holes down to low enough temperatures
to differentiate between this scaling behaviour, with a small scaling exponent, and
the non-vanishing constant entropy density of the solution that approaches $AdS_2\times\mathbb{R}^2$ in the IR.

Since the $AdS_2\times\mathbb{R}^2$ solution \eqref{ads2} is RG stable in the IR, 
the arguments presented in \cite{Donos:2012js} imply that
it is likely that there is some kind of bifurcating solution separating the metallic phase from the insulating
phase, for some specific value of $\lambda/\mu$ (at fixed $k/\mu$), and it would be interesting to identify it.

\subsubsection{$-1/3<\gamma\le -1/12$}
These models share some similarities with a model studied in \cite{Donos:2013eha}. 
In particular, the AdS-RN black hole is dynamically stable, but the $AdS_2\times\mathbb{R}^2$ solution
in the IR at $T=0$ is RG unstable at $k=0$, with relevant modes (i.e. $\delta_\phi<0$ at $k=0$ in \eqref{delphi}).
For Q-lattice deformations with small amplitude and small wavelength, i.e. $\lambda/\mu<<1$ and
$k/\mu>>1$, the black holes at $T=0$ should still approach the
$AdS_2\times\mathbb{R}^2$ solution \eqref{ads2} in the IR. 
This is because the lattice perturbation is expected to be irrelevant in the IR (i.e. $\delta_\phi>0$ in \eqref{delphi}).
These black holes will again be conventional holographic metals.
We have constructed such black holes for various values of $\lambda/\mu$, $k/\mu$, and 
$\lambda/\mu=1/10$, $k/\mu=3\sqrt{2}/4$ is a particular example.

As $k/\mu$ is reduced or $\lambda/\mu$ is increased, the lattice deformation can grow in the IR and eventually
become RG relevant inducing a transition
to a new IR behaviour at $T=0$. It is worth emphasising again that 
the RG instability of the $AdS_2\times\mathbb{R}^2$ IR region depends not just on $k$ but also on $v_{10}$, which depends
in a nonlinear way on the UV data. For $\gamma=-1/6$, $\lambda/\mu=1/10$ and $k/\mu=\sqrt{2}/20$
we have shown that the $T=0$ limit of the black holes do indeed approach the new
fixed point solutions in the IR. In particular we have checked that as $T\to 0$ the entropy of the black holes scales with temperature exactly as in 
\eqref{enscal}; see figure \ref{figone}(b). 
Notice from \eqref{conscal} that for the range $-1/3<\gamma\le -1/12$ the new ground states are insulators.
The black holes for these models provide a holographic example of a metal-insulator transition
which is driven by the metallic state becoming RG unstable.

\subsubsection{$-1/12<\gamma$}  
\begin{figure}
\centering
\hskip 1 em
\subfloat[]{\includegraphics[width=7.6cm]{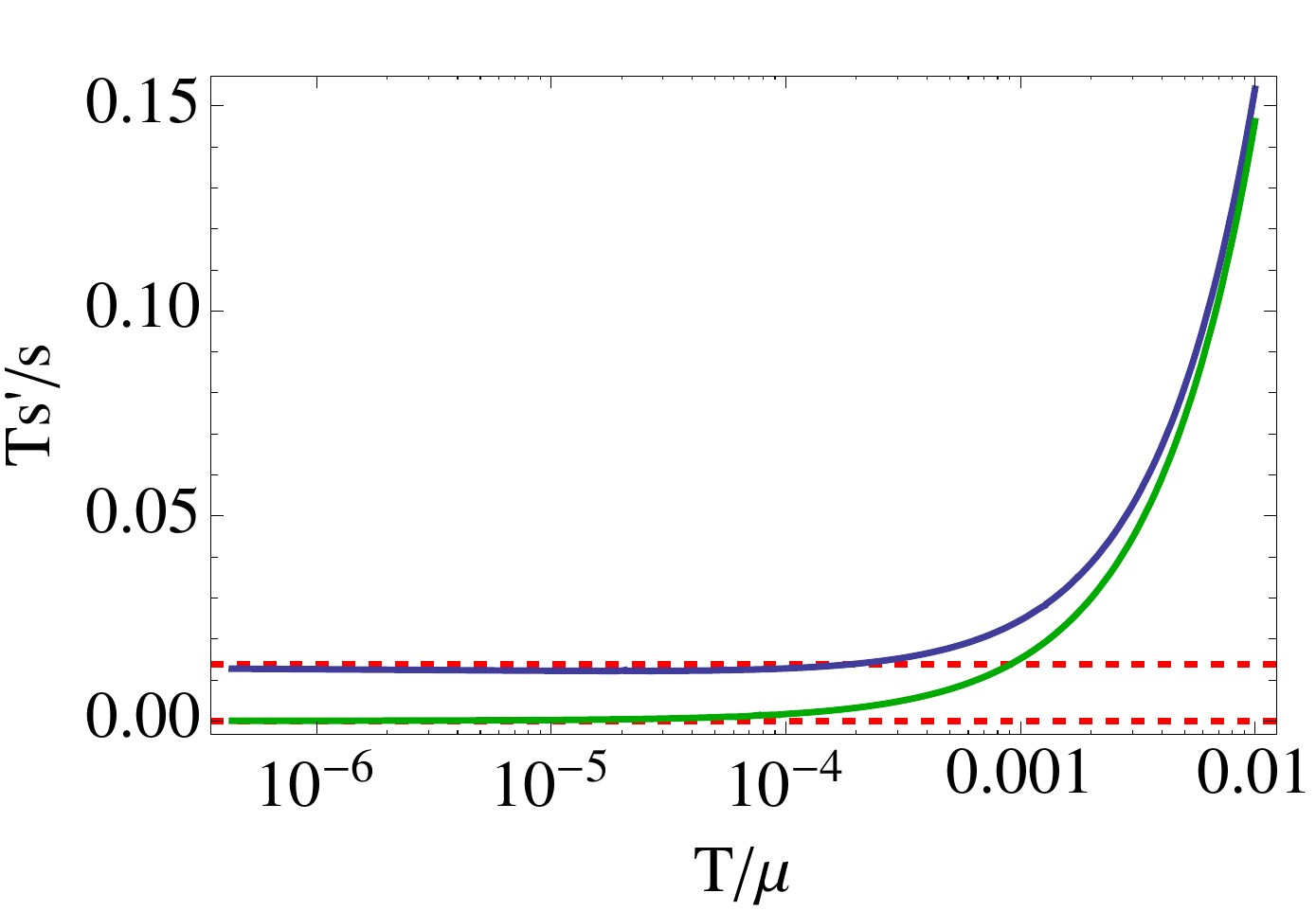}}
\subfloat[]{\includegraphics[width=7.3cm]{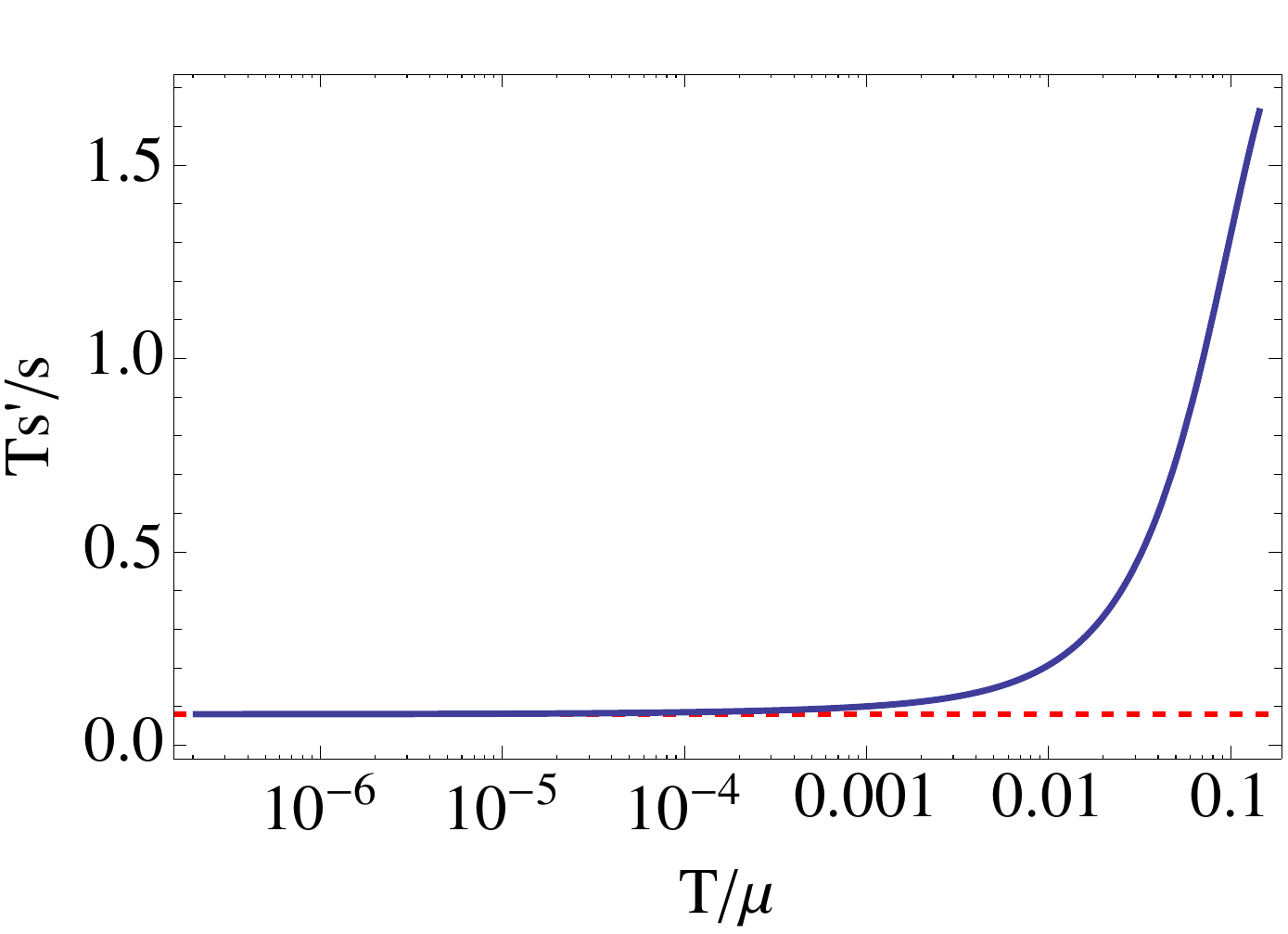}}\\
\subfloat[]{\includegraphics[width=7.3cm]{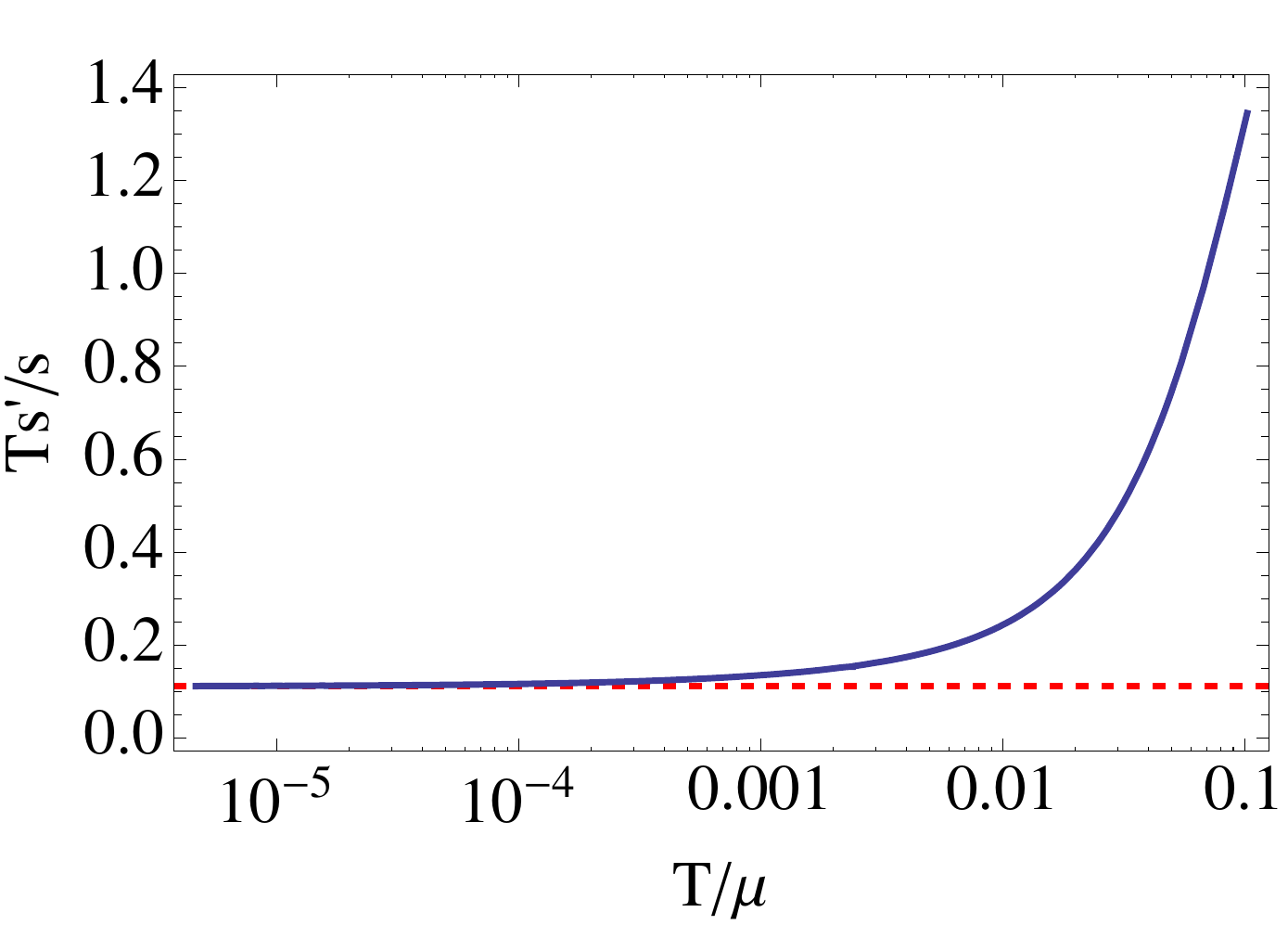}}
\subfloat[]{\includegraphics[width=7.3cm]{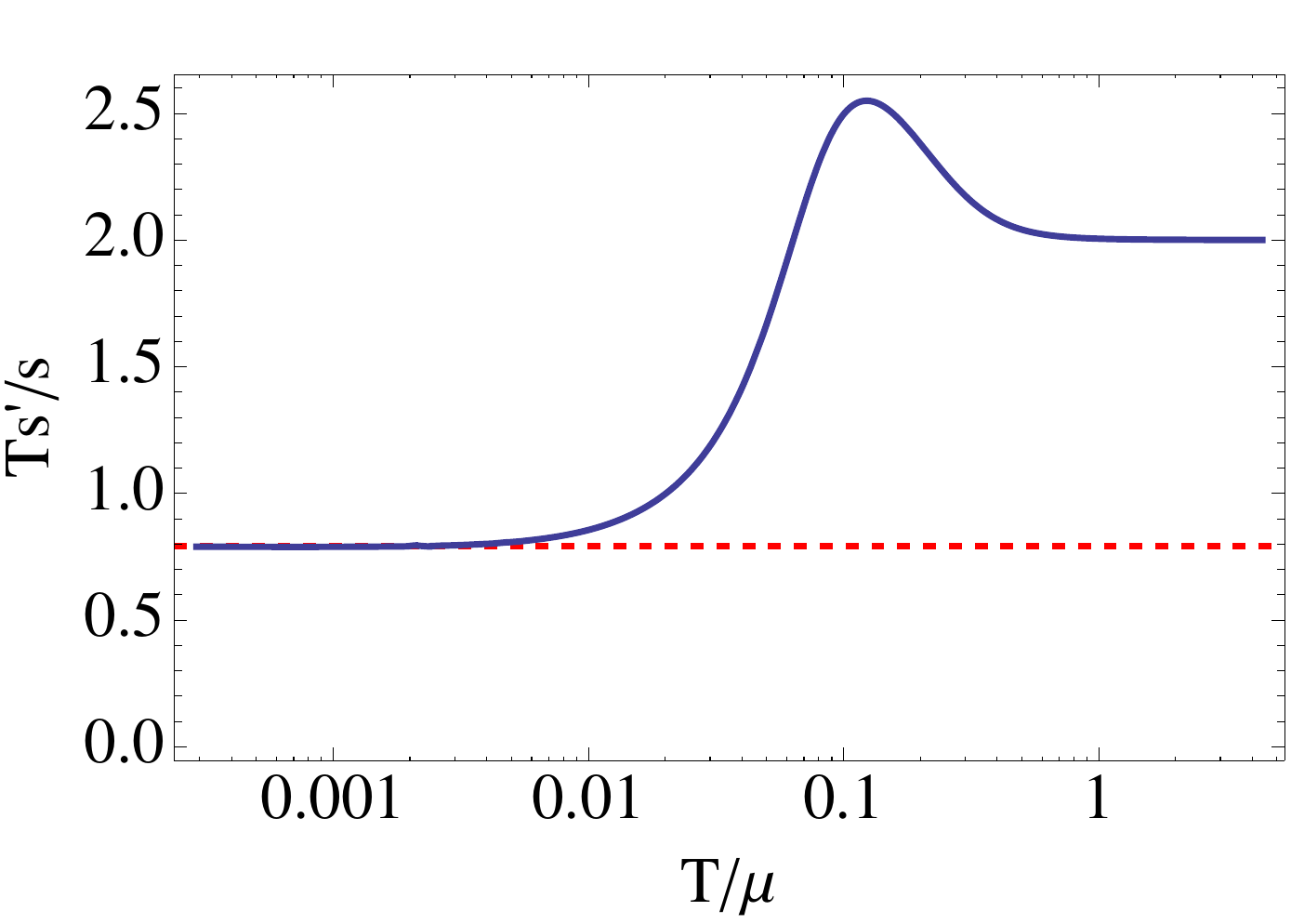}}\\
\caption{The blue curves indicate the
low-temperature behaviour of the entropy for the Q-lattice black holes 
which approach, in the IR and at $T=0$, the new 
fixed point solutions of section \ref{sec2}. The red-dashed
line indicate the scaling behaviour expected from \eqref{enscal}.
Panel (a) is for $\gamma=-2/3$, $k/\mu=\sqrt{2}/20$ and $\lambda/\mu=3/2$,
with $s\sim T^{0.014}$. The green curve in this panel shows another black hole solution 
for $\gamma=-2/3$, $k/\mu=\sqrt{2}/20$, but with a smaller lattice deformation of
$\lambda/\mu=1/10$, which approaches a translationally invariant $AdS_2\times\mathbb{R}^2$ solution
with non-zero entropy density.
Panel (b) is for 
$\gamma=-1/6$, $k/\mu=\sqrt{2}/20$ and $\lambda/\mu=1/10$ with $s\sim T^{0.080}$.
Panel (c) is for $\gamma=0$, $k/\mu=\sqrt{2}/20$ and $\lambda/\mu=1$ with $s\sim T^{0.11}$. 
Panel (d) is for $\gamma=9/2$, $k/\mu=\sqrt{2}/20$, $\lambda/\mu=1$ with $s\sim T^{0.79}$. 
}\label{figone}
\end{figure}

For these models the AdS-RN black hole is dynamically unstable. Thus, the AdS-RN black hole only describes the high temperature phase of the CFT
at finite $\mu$ when there is no Q-lattice. 
At some critical temperature, the scalar field $\phi$ will condense leading to a new class of black holes with scalar hair. While
it would be interesting to construct these, our aim here is to realise the new fixed point solutions found in the last section. We have found
for particular Q-lattices that this can indeed be achieved. For example, setting $\gamma=0$ we can take
$k/\mu=\sqrt{2}/20$ and $\lambda/\mu=1$ to find 
black holes whose zero temperature limit is the insulating ground states; see figure \ref{figone}(c).
We also considered $\gamma=9/2$ with 
$k/\mu=\sqrt{2}/20$, $\lambda/\mu=1$ and found black hole whose zero temperature limit is the new metallic ground states; see figure \ref{figone}(d).

\section{Isotropic ground states with no spatial translational symmetry}\label{sec4}
The holographic Q-lattices introduced in \cite{Donos:2013eha} and studied further here, provide a rich 
framework for studying metal-insulator
transitions and finding novel ground states. It is clear that there are many avenues for further investigation.
In this section we briefly discuss a new class of ground states 
which break translational symmetry in all spatial directions of the dual field theory. Moreover, we will see some
new features appearing in the conductivity.

We generalise the model \eqref{act2}, which has a single axion field to include two, $\chi_1$ and $\chi_2$:
\begin{align}\label{act3}
S=\int d^4 x\sqrt{-g}\left[R-\frac{c}{2}\left[(\partial\phi)^2+e^{2\phi}(\partial\chi_{1})^2+e^{2\phi}(\partial\chi_{2})^2\right]+n_1e^{\alpha\phi}-n_2\frac{e^{\gamma\phi}}{4}F^2\right]\, .
\end{align}
The scalar fields $\phi,\chi_1$ and $\chi_2$ can be viewed as parametrising three-dimensional hyperbolic 
space $H^3$. Following the discussion in section \ref{sec2}, the ground states with $\phi\to \infty$, which we will focus on,
can also arise in models where the scalar manifold locally approaches $H^3$. 
Alternatively, these models can be viewed as arising from set-ups where we have two complex scalar fields, 
with the full action having a $\mathbb{Z}_{2}$ symmetry which exchanges the two fields. 
In such set-ups we should choose the amplitudes of the two complex scalar fields
to be equal, and given by $\phi$,
with the two periodic axions $\chi_i$, having equal field strength squared. 
Finally, we note that we can easily scale our results to obtain solutions to models with no factors of $e^{2\phi}$ coupling to $(\partial \chi_i)^2$ as
we explain in the appendix.

With these comments in mind, we will
explore possible fixed point solutions for the action \eqref{act3} by making the ansatz
\begin{align}\label{ansatz2}
&ds^{2}=-U\,dt^{2}+U^{-1}\,dr^{2}+e^{2V}\,\left( dx_{1}^{2}+dx_{2}^{2}\right),\nn
&A=a\,dt,\nn
&\chi_{1}=k\,x_{1},\qquad \chi_{2}=k\,x_{2}\,,
\end{align}
with $U$, $V$, $a$ and $\phi$ being functions of $r$. Clearly such solutions will
break translation invariance in both of the spatial directions, but with an isotropic metric.
We again look for solutions with a power-law behaviour:
\begin{align}\label{gdstate2}
U=L^{-2}r^{u_1},\qquad e^{V}=e^{v_{0}}r^{v_{1}},\qquad a=a_0 r^{a_1},\qquad e^{\phi}=e^{\phi_0}r^{\phi_{1}}\,.
\end{align}
Substituting into the equations of motion arising from \eqref{act3}
we find a new class of solutions with
\begin{align}\label{sec41}
&a_{0}=0,\qquad e^{\left(2-\alpha\right)\,\phi_{0}}=\frac{n_{1}e^{2 v_{0}}\left(c-\alpha\left( \alpha-2\right) \right)}{k^{2}c\,\left(4+c-2 \alpha \right)}\,,\nn
&L^2=\frac{2e^{-\alpha\phi_{0}}}{n_{1}}\,\frac{(c+4-2\alpha)(c+(\alpha-2)(\alpha-6))}{\left(c+\left(\alpha-2 \right)^{2} \right)^{2}}\,,
\end{align}
and exponents
\begin{align}\label{sec42}
v_{1}=\frac{\left( \alpha-2\right)^{2}}{c+\left(\alpha-2\right)^{2}},\qquad \phi_{1}=\frac{2(2-\alpha)}{c+\left(\alpha-2\right)^{2}},\qquad u_{1}=\frac{2(4+c-2\alpha)}{c+\left(\alpha-2\right)^{2}}\,.
\end{align}
Observe that since $a_{0}=0$ these solutions do not carry any electric charge.
Also notice that for the particular choice of $\alpha=2$, and arbitrary $c>0,n_i$, we have neutral $AdS_{2}\times\mathbb{R}^{2}$ solutions 
that are supported by axions along both of the $\mathbb{R}^{2}$ directions:
\begin{align}\label{ads2explic}
&ds^{2}=L^2ds^2(AdS_2)+dx_{1}^{2}+dx_{2}^{2},\nn
&A=0,\quad \chi_{1}=k\,x_{1},\quad \chi_{2}=k\,x_{2}\,,
\end{align}
with $k^2=n_1/c$, $L^2=(2/n_1)e^{-2\phi_0}$ and $\phi=\phi_0$ a free parameter.

We now examine the static modes within the ansatz \eqref{ansatz2} about all of the new fixed point solutions. 
This is achieved by expanding the equations of motion around \eqref{gdstate2} via
\begin{align}\label{modeexp2}
U&=L^{-2}r^{u_1}(1+c_1 r^\delta),\quad e^{V}=e^{v_{0}}r^{v_{1}}(1+c_2r^\delta),\nn
 a&=c_3 \,r^{a_{1}+\delta},\qquad\qquad e^{\phi}=e^{\phi_0}r^{\phi_{1}}(1+c_4 r^\delta)\,,
\end{align}
for small $c_{i}$, and utilising the exponent shift for the gauge-field given by
\begin{align}
 a_{1}&=\frac{c+\left(\alpha-2\right)\left(\gamma-2\right)}{c+\left(\alpha-2\right)^{2}}\,.
 \end{align}
Substituting into the equations of motion and keeping terms linear in the $c_i$ then yields 
a single solution with $\delta=-1$, which corresponds to shifting the radial coordinate, and
three pairs of modes $\delta_{\pm}$ satisfying $\delta_{+}+\delta_{-}=1-u_{1}-2v_{1}$. One of these
has $\delta_{+}=0$ and the other two have
 \begin{align}\label{sol2deltas}
 \delta_{+}^{(1)}&=-\frac{c+\left(\alpha-6\right)\left(\alpha-2\right)}{2\left(c+\left(\alpha-2\right)^{2}\right)} +\frac{\sqrt{\left( c+\left( \alpha-6\right)\left(\alpha-2\right)\right)\,\left( 9c^{2}+16\left(\alpha-2 \right)^{2}\alpha-c\left(\alpha-2\right)\left( 7\alpha+22\right)\right)}}{2\sqrt{c}\left(c+\left(\alpha-2\right)^{2}\right)}\nn
 \delta_{+}^{(2)}&=-\frac{\left(\alpha-2\right)\left(\alpha-4-\gamma\right)}{c+\left(\alpha-2\right)^{2}}\,,
 \end{align}
(where here we have not assumed for these modes that $\delta_+>\delta_-$). 

The solution with $\delta_-=1-u_{1}-2v_{1}$ (associated with the $\delta_{+}=0$ mode)
 can be used to construct a small black hole, in an analogous manner to the discussion
in section \ref{sec:mode_an}. If $r_+$ is the location of the horizon the temperature is given by 
$T\propto r_+^{u_1-1}$ and the entropy density is given by $s\propto  r_+^{2v_1}$. It is natural to focus on the range of
parameters given by
\begin{align}\label{rangep}
2\leq \alpha<\sqrt{4+c}\,,
\end{align}
where the upper bound ensures that the black holes have zero surface gravity, $u_{1}>1$ at $T=0$, and the lower bound ensures that $\phi\rightarrow\infty$ as $T\to 0$. We can check that this implies $L^2,e^{\phi_0}>0$ as required.
Also note that for the small black holes $s\to 0$ as
$T\to 0$ except for the axionic $AdS_{2}\times\mathbb{R}^{2}$ solution \eqref{ads2explic}
with $\alpha=2$, where it goes to a non-zero constant,
as expected. It is also worth noting that the mode exponent $\delta_{+}^{(1)}$ in \eqref{sol2deltas} is always positive in the range \eqref{rangep}.

We noted above that the ground state solutions \eqref{gdstate2} do not carry any electric charge. 
However, there is a mode for switching on the electric field 
given by the constant $c_{3}$ in \eqref{modeexp2},
with $c_1=c_2=c_4=0$. Specifically, the mode we are interested in has $\delta_{+}^{(2)}$ in \eqref{sol2deltas} (observe that 
$\delta_{-}^{(2)}=-a_1$ and hence gives an uninteresting constant in \eqref{modeexp2}). Provided that
$\delta_{+}^{(2)}\ge 0$, which means\footnote{For $\gamma<\alpha-4$ there appear to be other fixed point solutions, differing from \eqref{gdstate2}, \eqref{sec41}, that are expansions (analogous to equation (10) of \cite{Donos:2012js}) about the hyper-scaling violating solutions \cite{Charmousis:2010zz}
which exist when $k=0$. When the $k=0$ solutions are realised as IR ground states arising from breaking translation invariance using an ionic or scalar
lattice (without axions), it has been argued that the DC resistivity would be exponentially suppressed at low temperatures \cite{Hartnoll:2012wm}. By contrast when axions are used
to break the translation invariance we expect that the $k\ne0$ ground states will give rise to power-law behaviour, with the exponent fixed by
the theory.}
\begin{align}\label{gammons}
\gamma\ge \alpha-4\,,
\end{align} 
or $\alpha=2$, 
this mode provides a mechanism for introducing charge in 
domain wall solutions interpolating between $AdS_4$ in the UV and fixed point solutions in the IR
in the context of more general classes of theories (analogous to the discussion in section \ref{sec3}).
Note in particular that when $\alpha=2$, corresponding to the new axionic $AdS_{2}\times\mathbb{R}^{2}$ solution, 
we have $\delta_{+}^{(2)}=0$ implying that the charge deformation will be a marginal deformation (as it is in the $AdS_{2}\times\mathbb{R}^{2}$ solution with translation symmetry arising in the standard AdS-RN black hole solution at $T=0$).
However, we note that since in \eqref{ads2explic} the wave number $k$ is fixed by specific constants in the model (i.e. $k^2=n_1/c$), it might be challenging to construct a domain wall solution that flows
from this fixed point in the IR to $AdS_4$ in the UV. It would be interesting to explore this further.

To investigate the optical conductivity associated with the new ground states we consider
the perturbation
\begin{align}
g_{tx_1}&=\delta h_{tx_1}(t,r)\,,\nn
A_{x_1}&={\delta a_{x_1}}(t,r)\,,\nn
\chi_1&=kx_1+\delta\chi_1(t,r)\,,
 \end{align}
around the background \eqref{gdstate2} (with no perturbation of $\chi_2$). 
Carrying out a similar calculation to that described 
in section \ref{sec:cond}, we find two decoupled modes involving the vector field perturbation.
By examining the small $\omega$ behaviour and using the matching procedure of \cite{Donos:2012ra}
we find that the scaling behaviour for the real part of the conductivity is given by
\begin{align}\label{conscal3}
\mathrm{Re}\sigma\propto \text{Max}(\omega^{{\left| 1-\frac{2(\alpha-2)\gamma}{4+c-\alpha^{2}}\right|}-1},\,\omega^{2+\frac{2\alpha(\alpha-2)}{4+c-\alpha^2} })\,,
\end{align}
where the absolute value in the first exponent, arising in the matching procedure, ensures that the smallest possible exponent is $-1$ when $\gamma=\frac{4+c-\alpha^2}{2(\alpha-2)}$.

Further insight can be obtained by calculating the DC conductivity at finite temperature. In section \ref{sec5} we show that the low temperature scaling is given by
\begin{align}\label{po}
\sigma_{DC}\sim T^{-\frac{2(\alpha-2)\gamma}{4+c-\alpha^2}}\,,
\end{align}
which is aligned with the scaling of the optical conductivity provided that the first exponent in \eqref{conscal3} is the dominate one and that
$ 1-\frac{2(\alpha-2)\gamma}{4+c-\alpha^{2}}\ge 0$. 
Assuming the constraints on the parameters given in \eqref{rangep}, \eqref{gammons}, by considering the $\omega\to 0$ limit of $\mathrm{Re}\sigma$ from \eqref{conscal3} and
the $T\to 0$ limit of $\sigma_{DC}$ form \eqref{po},
there are three broad cases, which include various novel metallic
and insulating behaviour:

Firstly, when $\alpha-4\le \gamma\le\frac{4+c-\alpha^2}{2(\alpha-2)}$, the scaling of $\mathrm{Re}\sigma(\omega)$ is given by the first exponent 
in \eqref{conscal3} and is aligned with that of $\sigma_{DC}$. These cases include insulators with $\sigma_{DC}=0$ when $\alpha-4\le\gamma<0$,
metals with $\sigma_{DC}$ a nonzero constant when $\gamma=0$ and metals with infinite $\sigma_{DC}$ when 
$0<\gamma\le \frac{4+c-\alpha^2}{2(\alpha-2)}$.

Secondly, when $\frac{4+c-\alpha^2}{2(\alpha-2)}<\gamma<1+\frac{2(4+c-\alpha^2)}{(\alpha-2)}$, the scaling of $\mathrm{Re}\sigma(\omega)$
is given by the first exponent in \eqref{conscal3} and is not aligned with that of $\sigma_{DC}$. These cases include metals 
when $\frac{4+c-\alpha^2}{2(\alpha-2)}<\gamma<\frac{4+c-\alpha^2}{(\alpha-2)}$ where $\mathrm{Re}\sigma(\omega\to 0)$ and $\sigma_{DC}$ both diverge.
It includes metals when $\gamma=\frac{4+c-\alpha^2}{(\alpha-2)}$ where $\mathrm{Re}\sigma(\omega\to 0)$ seems to be a constant and yet
$\sigma_{DC}$ diverges like $T^{-2}$, which indicates that $\mathrm{Re}\sigma(\omega)$ has a delta function. Finally, when
$\frac{4+c-\alpha^2}{(\alpha-2)}<\gamma<1+\frac{2(4+c-\alpha^2)}{(\alpha-2)}$ we seem to have $\mathrm{Re}\sigma(\omega\to 0)$ is vanishing but 
$\sigma_{DC}$ is diverging, again indicating that $\mathrm{Re}\sigma(\omega)$ has a delta function.

Thirdly, when $1+\frac{2(4+c-\alpha^2)}{(\alpha-2)}<\gamma$ we have the scaling of $\mathrm{Re}\sigma(\omega)$ is now given by the second exponent in
\eqref{conscal3}. We seem to have $\mathrm{Re}\sigma(\omega\to 0)$ is vanishing  but a diverging $\sigma_{DC}$ and hence there seems to be
a delta function in
$\mathrm{Re}\sigma(\omega)$.

For the second and third cases, there is a misalignment of the scaling of the $DC$ and the optical conductivities, which indicates that there
is a parametrical separation of scales between the thermal time scale and the momentum relaxation time scale. It would certainly be interesting to further understand the transport properties of these new metallic phases. 

The delta functions in the optical conductivity at zero temperature that we are inferring above, are associated with
different physics to the delta functions that appear when a holographic lattice becomes an irrelevant deformation of a translationally invariant 
$AdS_2\times\mathbb{R}^2$ solution
in the IR, as in \cite{Hartnoll:2012rj}. In particular, in the latter examples, the weight of the delta function is proportional to the electric charge of the domain wall solution, unlike the present situation. 
This can be seen by considering electrically neutral domain wall solutions that approach our new fixed point solutions in the IR, where we
would be lead to similar conclusions concerning the scaling and hence the existence of delta functions. We expect this to be a more general phenomenon for geometries with a running background scalar. 

The different scaling properties of the DC and optical conductivity that we have seen is reminiscent of the cuprates. Recall that for the cuprates
$\sigma_{DC}\sim T^{-1}$ and for $\omega/T\gtrsim 1.5$ there is a scaling behaviour of the form $|\sigma|\sim \omega^{-0.65}$ \cite{2003Natur.425..271M}. It is worth emphasising, however, that
our holographic results concerning the scaling of the optical conductivity, which was derived at $T=0$, is expected to
hold for $T<<\omega$ (and $\omega<<\mu$).

Finally,  we also highlight that for $\alpha=2$, corresponding to the axionic $AdS_2\times\mathbb{R}^2$ solution
\eqref{ads2explic}, the DC conductivity will be constant; this is to be contrasted to the delta-function
that appears in the DC conductivity for the standard, translationally  invariant
$AdS_2\times\mathbb{R}^2$ solution.

\section{Calculating the DC conductivity}\label{sec5}

In this section we present a new approach to calculating the DC conductivity for black hole solutions and use it to obtain analytic 
expressions for the black hole solutions relevant to the previous sections. We will illustrate the approach using 
the following general class of models
\begin{align}\label{eq:aniso_model}
S=\int d^4 x\sqrt{-g}\left[R-\frac{c}{2}\left[(\partial\phi)^2+\Phi_1(\phi)(\partial\chi_{1})^2+\Phi_2(\phi)(\partial\chi_{2})^2\right]+V(\phi)-\frac{Z(\phi)}{4}F^2\right]
\, ,
\end{align}
which involves four functions, $\Phi_i,V$ and $Z$, of the real scalar field $\phi$. 
We will assume the model admits a unit radius $AdS_4$ vacuum
with $\phi=0$ and take $\Phi_i(0), Z(0)$ to be constants greater than or equal to zero.
We assume that in this $AdS_4$ vacuum the field $\phi$ is dual to a relevant operator with dimension $\Delta<3$.

The background geometry is assumed to be of the form
\begin{align}\label{ansatz3}
&ds^{2}=-U\,dt^{2}+U^{-1}\,dr^{2}+e^{2V_1}dx_{1}^{2}+e^{2V_2}dx_{2}^{2},\nn
&A=a\,dt,\qquad \chi_{1}=k\,x_{1},\qquad \chi_{2}=k\,x_{2}\,,
\end{align}
where $U,V_i,a$ and $\phi$ are functions of $r$ only.
As $r\to\infty$ the metric approaches the $AdS_4$ vacuum with $a\sim\mu+q/r+...$ and the asymptotic behaviour of
$\phi$ corresponds to a Q-lattice deformation with, in particular $\phi\to 0$ with a rate dependent on the dimension of the dual operator
($\phi\sim\lambda r^{\Delta-3}$ in the standard quantisation). 
We also assume that there
is a regular event horizon at $r=r_+$ where $V_i$ and $\phi$ approach constant values, while $a\sim a_+(r-r_+)$ and $U\sim4\pi T(r-r_+)$.
Below we will use ingoing Eddington-Finklestein coordinates $(v,r)$ where $v=t+\frac{1}{4\pi T}\ln(r-r_+)$.
It is useful to solve the gauge-field equations of motion exactly via
\begin{align}
q=-e^{V_1+V_2}Z(\phi)a'\,.
\end{align}

The usual approach to calculating the DC conductivity is to switch on an AC electric field with frequency $\omega$, find the response
and then take the $\omega\to 0$ limit. Instead we will switch on a constant electric field from the start. Specifically we consider
the following perturbation
\begin{align}\label{dcxan}
A_{x_1}&=-Et +{\delta a_{x_1}}(r)\,,\nn
g_{tx_1}&=e^{{2V_1}}\delta h_{tx_1}(r)\,,\nn
g_{rx_1}&=e^{{2V_1}}\delta h_{rx_1}(r)\,,\nn
\chi_1&=kx_1+\delta\chi_1(r)\,,
 \end{align}
which one can check is consistent with the linearised equations of motion. We find one equation which we can algebraically 
solve for $\delta h_{rx_1}$ giving
\begin{align}\label{gravcon}
\delta h_{rx_1}=\frac{Eqe^{-V_1-V_2}}{k^2\Phi_1(\phi)U}+\frac{\delta\chi_1'}{k},
\end{align}
(which requires that $k\ne 0$)
and two further independent differential equations. The first is given by
\begin{align}\label{heq}
\delta h_{tx_1}''+(3{V_1}' +{V_2}')\delta h_{tx_1}'-\frac{      e^{-2 {V_1}} {k}^2 {\Phi_1}(\phi) }{U}{\delta h_{tx_1}}-e^{-3 {V_1}-{V_2}} q \delta{a_{x_1}'}  =0
\end{align}
while the second, arising from the gauge-field equations of motion, can be integrated once to obtain the constraint
\begin{align}\label{jexp}
j=-e^{{V_2}-{V_1} }Z(\phi) U \delta a_{x_1}'+q \delta h_{tx_1}
\end{align}
where $j$ is a constant.

To make further progress we need to impose boundary conditions at infinity and at the black hole horizon. It is important to note that
$\delta\chi_1$ only appears in \eqref{gravcon}; we will assume that $\delta\chi_1$ is analytic at the black hole event horizon and falls off 
sufficiently fast at infinity.
Regularity of the perturbation as $r\to r_+$ is obtained by switching to Eddington-Finklestein coordinates. 
The gauge-field will be well defined if we demand that $\delta{a_{x_1}}\sim-\frac{E}{4\pi T}\ln(r-r_+)+{\cal O}(r-r_+)$ and hence $\delta a_{x_1}'\sim-\frac{E}{U}+{\cal O}(r-r_+)$.
For the metric perturbation, we see from \eqref{gravcon} that $\delta h_{rx_1}$ is diverging; this can be remedied by demanding that
$\delta h_{tx_1}$ behaves as $\delta h_{tx_1}\sim \frac{Eqe^{-V_1-V_2}}{k^2\Phi_1(\phi)}|_{r=r_+}+{\cal O}(r-r_+)$. Notice that the behaviour for 
$\delta a_{x_1}'$ and $\delta h_{tx_1}$ is consistent with \eqref{heq}.

We next consider the behaviour as $r\to\infty$. From the expression for the gauge-field we see that we have a deformation of an electric field
in the $x_1$ direction with strength $E$. Now, \eqref{heq} has two independent solutions, one of which behave as $r^0$ and the other as $r^{-3}$; in order to have no additional deformations we demand that the coefficient of the former vanishes.
Observe that since we have also specified the boundary condition of
$\delta h_{tx_1}$ at the horizon this completely specifies the solution of \eqref{heq}. In addition, notice
from \eqref{gravcon} 
that
for suitable $\delta\chi_1$ the fall-off of $\delta h_{rx_1}$ 
can be as weak as desired and will not affect the boundary data\footnote{Note that one can attempt to work from the start in a radial gauge with $\delta h_{rx_1}=0$. After solving \eqref{gravcon} one concludes that one needs to modify
the ansatz in \eqref{dcxan} to include a term $ct$ in $\chi_1$, where $c$ is a suitable constant, 
to ensure that $d\chi_1$ is well defined at the event horizon. This in turn leads to an extra term in \eqref{heq} and hence to a
deformation at the boundary $\delta h_{tx_1}\sim c/k$. Following the discussion at the end of section \ref{sec3} of \cite{Donos:2013eha} we can deal with this by carrying out a
diffeomorphism $x_1\to x_1-c/k t+f(r)$ where the function $f(r)$ satisfies $\left. \partial_{r}f \right|_{r=r_{+}}=-\frac{1}{U}\frac{c}{k}$,
is smooth in the bulk and falls off sufficiently fast at the boundary. This procedure also leads to \eqref{dcresult}.}.
From \eqref{jexp} we can now conclude that we should take
$\delta a_{x_1}\sim \frac{j}{r}$ where we now see that $j$ is
the electric current in the $x_1$ direction which is induced by the electric field deformation.

The DC conductivity is thus given by $\sigma_{DC}=j/E$. To obtain our final result we now simply 
evaluate the right hand side of \eqref{jexp} at the black hole event horizon $r=r_+$. Doing so we obtain
the elegant expression
\begin{align}\label{dcresult}
\sigma_{DC}=\left[e^{-{V_1}+{V_2}} Z(\phi)+\frac{ q^2e^{-{V_1}-{V_2}}}{{k}^2 {\Phi_1}(\phi)}\right]_{r=r_+}\,,
\end{align}
which gives the DC conductivity of the black holes, for all temperatures, in terms of black hole horizon data.

\subsection{Examples}
The expression \eqref{dcresult}
can be used to obtain the DC conductivity for the black holes constructed in section \ref{sec3} for all temperatures. The low temperature behaviour, as the black holes
approach the fixed point solutions of section \ref{sec3}, can be obtained by using the small black hole solutions constructed in section
\ref{sec:mode_an}. We find that both terms in \eqref{dcresult} scale in the same way and we obtain $\sigma_{DC}\sim T^{\frac{(1+\gamma)(3-\gamma)}{9+2\gamma+\gamma^2}}$. Similarly, for the conventional holographic metals described by black holes 
approaching $AdS_2\times\mathbb{R}^2$ in the IR, we find that the second term in \eqref{dcresult} dominates and we 
get $\sigma_{DC}\sim T^{-2\delta_\phi}$, with $\delta_\phi$ given in \eqref{delphi}, 
recovering the
result of \cite{Hartnoll:2012rj} using the memory matrix formalism.

For the electrically neutral ground states discussed in section \ref{sec4}, by similarly analysing the small black hole solutions, 
we find that the contribution from the second term
in \eqref{dcresult} is sub-dominant to the first and that the first term leads to the scaling behaviour
$\sigma_{DC}\sim T^{-\frac{2(\alpha-2)\gamma}{4+c-\alpha^2}}$.

We can also check the formula for $\sigma_{DC}$ with some other examples in the literature. Black holes with holographic Q-lattices were
constructed numerically in \cite{Donos:2013eha} and the DC conductivity was obtained for a range of finite temperatures by 
taking the $\omega\to 0$ limit of the optical conductivity. We can compare these results (given in figures 1(c) and 2(a) of \cite{Donos:2013eha})
to the results obtained by substituting the horizon data of the background black hole solutions into \eqref{dcresult} and we
find excellent agreement.
We have also checked that the formula \eqref{dcresult} gives the analytic expression for the constant DC conductivity found for the 
analytic black holes found in \cite{Andrade:2013gsa}. Finally, we notice that second term in \eqref{dcresult} is singular as $k\to 0$. However, 
if we consider neutral, translationally invariant black holes with $q=k=\chi=0$, we can set the metric 
perturbation to zero in \eqref{dcxan} and we are lead to $\sigma_{DC}$ given by the first term in \eqref{dcresult}. A simple example is the
AdS-Schwarzschild black hole with $\phi=0$ for which we obtain the known result $\sigma_{DC}=Z(0)$ for all temperatures.
\vskip .8 cm
{\bf Note added:} Results with some overlap with this paper appear in \cite{Gouteraux:2014hca}.

\section*{Acknowledgements}
We thank Blaise Gout\'eraux, Elias Kiritsis, David Tong and especially Mike Blake for helpful discussions. 
The work is supported by STFC grant ST/J0003533/1 and also by the European Research Council under the European Union's Seventh Framework Programme (FP7/2007-2013), ERC Grant agreement STG 279943 and ERC Grant agreement ADG 339140.

\appendix
\section{Non-translationally invariant $\eta$-geometries}
We briefly comment on fixed point solutions of theories with massless axions of the form
\begin{align}\label{appa}
S=\int d^4 x\sqrt{-g}\left[R-\frac{c}{2}\left[(\partial\phi)^2+(\partial\chi_{1})^2+(\partial\chi_{2})^2\right]+n_1e^{\alpha\phi}-n_2\frac{e^{\gamma\phi}}{4}F^2\right]\, .
\end{align}
In fact we can obtain the solutions by taking a scaling limit of the solutions that we found in section \ref{sec4}. Specifically
we scale the fields via $\phi\to\beta\phi,\chi\to\beta\chi$ and the parameters via 
$c\to c/\beta^2,\alpha\to\alpha/\beta,\gamma\to\gamma/\beta$ and then set $\beta=0$. 
From the fixed point solutions \eqref{sec41}, \eqref{sec42}, after scaling $k\to \beta k,\phi_0\to\beta\phi_0$,
we find a new class of neutral solutions for the action \eqref{appa} given by the ansatz \eqref{ansatz2} with
\begin{align}
&a_{0}=0,\qquad e^{-\alpha\,\phi_{0}}=\frac{n_{1}e^{2 v_{0}}\left(c-\alpha^2 \right)}{k^{2}c^2},\qquad
L^2=\frac{2e^{-\alpha\phi_{0}}}{n_{1}}\,\frac{c}{c+\alpha^2}\,,
\end{align}
and exponents
\begin{align}
v_{1}=\frac{ \alpha^2}{c+\alpha^{2}},\qquad \phi_{1}=\frac{-2\alpha}{c+\alpha^{2}},\qquad u_{1}=\frac{2c}{c+\alpha^{2}}\,.
\end{align}

It is illuminating to carry out the coordinate transformation $r=\rho^\frac{1}{1-u_1}$. In particular, the metric reads
\begin{align}
ds^2=\frac{1}{\rho^\eta}\left(-L^{-2}\frac{dt^2}{\rho^2}+\frac{L^2}{(1-u_1)^2}\frac{\rho^2}{\rho^2}+e^{2v_0}(dx_1^2+dx_2^2)\right)\,,
\end{align}
with $\eta=\frac{2-u_1}{u_1-1}=\frac{2\alpha^2}{c-\alpha^2}$. This is a so-called ``$\eta$-geometry"  i.e. a geometry 
conformal to $AdS_2\times\mathbb{R}^2$
which has been discussed as a framework for holographically studying semi-local quantum criticality \cite{Hartnoll:2012wm,Donos:2012yi} 
(earlier work appeared in \cite{Gubser:2009qt}). 
The perturbations of these new axionic $\eta$-geometries, including
the two-point functions, can be obtained by scaling the results of section \ref{sec4}. 
Note that the special case of $\alpha=0$ the axionic $AdS_2\times\mathbb{R}^2$
solution is the same as that in equation (2.10) of \cite{Andrade:2013gsa} after setting $\mu=0$. 
In contrast to \cite{Andrade:2013gsa}, these solutions can be viewed,
in the context of the discussion at the beginning of section \ref{sec4},
as IR ground states of a holographic lattice with, for example, the functions $\Phi_{1},\Phi_{2}$ in \eqref{eq:aniso_model} 
approaching constants in the IR and the UV.

Finally, we note that the analysis of this appendix can also be applied to the solutions of section \ref{sec2} to obtain new anisotropic
fixed point solutions.
 
\appendix
\bibliographystyle{utphys}
\bibliography{helical}{}

\providecommand{\href}[2]{#2}\begingroup\raggedright\begin{thebibliography}{10}

\bibitem{edwards}
P.~Edwards, R.~L. Johnston, C.~N.~R. Rao, D.~P. Tunstall, and F.~Hensel, ``{The
  metal insulator transition: a perspective},''
  \href{http://dx.doi.org/10.1098/rsta.1998.0146}{{\em Phil.Trans.Roy.Soc.Lond}
  {\bfseries A356} (1998) 5--22}.

\bibitem{gebhard}
F.~Gebhard, {\em {The Mott Metal-Insulator Transition}}.
\newblock Springer, Berlin, 2000.

\bibitem{Dobrosavljevic}
V.~Dobrosavljevic, ``{Introduction to Metal-Insulator Transitions},''
\href{http://arxiv.org/abs/1112.6166}{{\ttfamily arXiv:1112.6166
  [cond-mat.str-el]}}.

\bibitem{edwards2}
P.~Edwards, M.~Lodge, F.~Hensel, and R.~Redmer, ``{...a metal conducts and a
  non-metal doesn't},'' \href{http://dx.doi.org/10.1098/rsta.2009.0282}{{\em
  Phil.Trans.Roy.Soc.Lond} {\bfseries A368} (2010) 941--965}.

\bibitem{sachdev}
S.~Sachdev, {\em {Quantum Phase Transitions}}.
\newblock Cambridge University Press, 2011.

\bibitem{Iqbal:2011in}
N.~Iqbal, H.~Liu, and M.~Mezei, ``{Semi-local quantum liquids},''
  \href{http://dx.doi.org/10.1007/JHEP04(2012)086}{{\em JHEP} {\bfseries 1204}
  (2012) 086},
\href{http://arxiv.org/abs/1105.4621}{{\ttfamily arXiv:1105.4621 [hep-th]}}.

\bibitem{Nakamura:2009tf}
S.~Nakamura, H.~Ooguri, and C.-S. Park, ``{Gravity Dual of Spatially Modulated
  Phase},'' \href{http://dx.doi.org/10.1103/PhysRevD.81.044018}{{\em Phys.
  Rev.} {\bfseries D81} (2010) 044018},
\href{http://arxiv.org/abs/0911.0679}{{\ttfamily arXiv:0911.0679 [hep-th]}}.

\bibitem{Donos:2011bh}
A.~Donos and J.~P. Gauntlett, ``{Holographic striped phases},''
  \href{http://dx.doi.org/10.1007/JHEP08(2011)140}{{\em JHEP} {\bfseries 1108}
  (2011) 140},
\href{http://arxiv.org/abs/1106.2004}{{\ttfamily arXiv:1106.2004 [hep-th]}}.

\bibitem{Bergman:2011rf}
O.~Bergman, N.~Jokela, G.~Lifschytz, and M.~Lippert, ``{Striped instability of
  a holographic Fermi-like liquid},''
  \href{http://dx.doi.org/10.1007/JHEP10(2011)034}{{\em JHEP} {\bfseries 10}
  (2011) 034},
\href{http://arxiv.org/abs/1106.3883}{{\ttfamily arXiv:1106.3883 [hep-th]}}.

\bibitem{Vegh:2013sk}
D.~Vegh, ``{Holography without translational symmetry},''
\href{http://arxiv.org/abs/1301.0537}{{\ttfamily arXiv:1301.0537 [hep-th]}}.

\bibitem{Davison:2013jba}
R.~A. Davison, ``{Momentum relaxation in holographic massive gravity},''
  \href{http://dx.doi.org/10.1103/PhysRevD.88.086003}{{\em Phys.Rev.}
  {\bfseries D88} (2013) 086003},
\href{http://arxiv.org/abs/1306.5792}{{\ttfamily arXiv:1306.5792 [hep-th]}}.

\bibitem{Blake:2013bqa}
M.~Blake and D.~Tong, ``{Universal Resistivity from Holographic Massive
  Gravity},'' \href{http://dx.doi.org/10.1103/PhysRevD.88.106004}{{\em
  Phys.Rev.} {\bfseries D88} (2013) 106004},
\href{http://arxiv.org/abs/1308.4970}{{\ttfamily arXiv:1308.4970 [hep-th]}}.

\bibitem{Davison:2013txa}
R.~A. Davison, K.~Schalm, and J.~Zaanen, ``{Holographic duality and the
  resistivity of strange metals},''
\href{http://arxiv.org/abs/1311.2451}{{\ttfamily arXiv:1311.2451 [hep-th]}}.

\bibitem{Karch:2007pd}
A.~Karch and A.~O'Bannon, ``{Metallic AdS/CFT},''
  \href{http://dx.doi.org/10.1088/1126-6708/2007/09/024}{{\em JHEP} {\bfseries
  0709} (2007) 024},
\href{http://arxiv.org/abs/0705.3870}{{\ttfamily arXiv:0705.3870 [hep-th]}}.

\bibitem{Faulkner:2010zz}
T.~Faulkner, N.~Iqbal, H.~Liu, J.~McGreevy, and D.~Vegh, ``{Strange metal
  transport realized by gauge/gravity duality},''
\href{http://dx.doi.org/10.1126/science.1189134}{{\em Science} {\bfseries 329}
  (2010) 1043--1047}.

\bibitem{Hartnoll:2009ns}
S.~A. Hartnoll, J.~Polchinski, E.~Silverstein, and D.~Tong, ``{Towards strange
  metallic holography},'' \href{http://dx.doi.org/10.1007/JHEP04(2010)120}{{\em
  JHEP} {\bfseries 1004} (2010) 120},
\href{http://arxiv.org/abs/0912.1061}{{\ttfamily arXiv:0912.1061 [hep-th]}}.

\bibitem{Hartnoll:2012rj}
S.~A. Hartnoll and D.~M. Hofman, ``{Locally Critical Resistivities from Umklapp
  Scattering},'' \href{http://dx.doi.org/10.1103/PhysRevLett.108.241601}{{\em
  Phys.Rev.Lett.} {\bfseries 108} (2012) 241601},
\href{http://arxiv.org/abs/1201.3917}{{\ttfamily arXiv:1201.3917 [hep-th]}}.

\bibitem{Horowitz:2012ky}
G.~T. Horowitz, J.~E. Santos, and D.~Tong, ``{Optical Conductivity with
  Holographic Lattices},''
  \href{http://dx.doi.org/10.1007/JHEP07(2012)168}{{\em JHEP} {\bfseries 1207}
  (2012) 168},
\href{http://arxiv.org/abs/1204.0519}{{\ttfamily arXiv:1204.0519 [hep-th]}}.

\bibitem{Horowitz:2012gs}
G.~T. Horowitz, J.~E. Santos, and D.~Tong, ``{Further Evidence for
  Lattice-Induced Scaling},''
  \href{http://dx.doi.org/10.1007/JHEP11(2012)102}{{\em JHEP} {\bfseries 1211}
  (2012) 102},
\href{http://arxiv.org/abs/1209.1098}{{\ttfamily arXiv:1209.1098 [hep-th]}}.

\bibitem{Horowitz:2013jaa}
G.~T. Horowitz and J.~E. Santos, ``{General Relativity and the Cuprates},''
\href{http://arxiv.org/abs/1302.6586}{{\ttfamily arXiv:1302.6586 [hep-th]}}.

\bibitem{Donos:2012js}
A.~Donos and S.~A. Hartnoll, ``{Interaction-driven localization in
  holography},'' \href{http://dx.doi.org/10.1038/nphys2701}{{\em Nature Phys.}
  {\bfseries 9} (2013) 649--655},
\href{http://arxiv.org/abs/1212.2998}{{\ttfamily arXiv:1212.2998}}.

\bibitem{Ling:2013nxa}
Y.~Ling, C.~Niu, J.-P. Wu, and Z.-Y. Xian, ``{Holographic Lattice in
  Einstein-Maxwell-Dilaton Gravity},''
  \href{http://dx.doi.org/10.1007/JHEP11(2013)006}{{\em JHEP} {\bfseries 1311}
  (2013) 006},
\href{http://arxiv.org/abs/1309.4580}{{\ttfamily arXiv:1309.4580 [hep-th]}}.

\bibitem{Chesler:2013qla}
P.~Chesler, A.~Lucas, and S.~Sachdev, ``{Conformal field theories in a periodic
  potential: results from holography and field theory},''
  \href{http://dx.doi.org/10.1103/PhysRevD.89.026005}{{\em Phys.Rev.}
  {\bfseries D89} (2014) 026005},
\href{http://arxiv.org/abs/1308.0329}{{\ttfamily arXiv:1308.0329 [hep-th]}}.

\bibitem{Donos:2013eha}
A.~Donos and J.~P. Gauntlett, ``{Holographic Q-lattices},''
\href{http://arxiv.org/abs/1311.3292}{{\ttfamily arXiv:1311.3292 [hep-th]}}.

\bibitem{Andrade:2013gsa}
T.~Andrade and B.~Withers, ``{A simple holographic model of momentum
  relaxation},''
\href{http://arxiv.org/abs/1311.5157}{{\ttfamily arXiv:1311.5157 [hep-th]}}.

\bibitem{Bardoux:2012aw}
Y.~Bardoux, M.~M. Caldarelli, and C.~Charmousis, ``{Shaping black holes with
  free fields},'' \href{http://dx.doi.org/10.1007/JHEP05(2012)054}{{\em JHEP}
  {\bfseries 1205} (2012) 054},
\href{http://arxiv.org/abs/1202.4458}{{\ttfamily arXiv:1202.4458 [hep-th]}}.

\bibitem{Iizuka:2012iv}
N.~Iizuka, S.~Kachru, N.~Kundu, P.~Narayan, N.~Sircar, {\em et al.}, ``{Bianchi
  Attractors: A Classification of Extremal Black Brane Geometries},''
  \href{http://dx.doi.org/10.1007/JHEP07(2012)193}{{\em JHEP} {\bfseries 1207}
  (2012) 193},
\href{http://arxiv.org/abs/1201.4861}{{\ttfamily arXiv:1201.4861 [hep-th]}}.

\bibitem{Donos:2012gg}
A.~Donos and J.~P. Gauntlett, ``{Helical superconducting black holes},''
  \href{http://dx.doi.org/10.1103/PhysRevLett.108.211601}{{\em Phys.Rev.Lett.}
  {\bfseries 108} (2012) 211601},
\href{http://arxiv.org/abs/1203.0533}{{\ttfamily arXiv:1203.0533 [hep-th]}}.

\bibitem{Iizuka:2012pn}
N.~Iizuka, S.~Kachru, N.~Kundu, P.~Narayan, N.~Sircar, {\em et al.},
  ``{Extremal Horizons with Reduced Symmetry: Hyperscaling Violation, Stripes,
  and a Classification for the Homogeneous Case},''
  \href{http://dx.doi.org/10.1007/JHEP03(2013)126}{{\em JHEP} {\bfseries 1303}
  (2013) 126},
\href{http://arxiv.org/abs/1212.1948}{{\ttfamily arXiv:1212.1948}}.

\bibitem{Donos:2013woa}
A.~Donos, J.~P. Gauntlett, and C.~Pantelidou, ``{Competing p-wave orders},''
  \href{http://dx.doi.org/10.1088/0264-9381/31/5/055007}{{\em
  Class.Quant.Grav.} {\bfseries 31} (2014) 055007},
\href{http://arxiv.org/abs/1310.5741}{{\ttfamily arXiv:1310.5741 [hep-th]}}.

\bibitem{Donos:2012ra}
A.~Donos and S.~A. Hartnoll, ``{Universal linear in temperature resistivity
  from black hole superradiance},''
  \href{http://dx.doi.org/10.1103/PhysRevD.86.124046}{{\em Phys.Rev.}
  {\bfseries D86} (2012) 124046},
\href{http://arxiv.org/abs/1208.4102}{{\ttfamily arXiv:1208.4102 [hep-th]}}.

\bibitem{toapp}
A.~Donos, B.~Gouteraux, and E.~Kiritsis {\em {Work in progress and talk by E.
  Kiritsis at the Isaac Newton Institute, Sept. 2013}} .

\bibitem{Hartnoll:2012wm}
S.~A. Hartnoll and E.~Shaghoulian, ``{Spectral weight in holographic scaling
  geometries},'' \href{http://dx.doi.org/10.1007/JHEP07(2012)078}{{\em JHEP}
  {\bfseries 1207} (2012) 078},
\href{http://arxiv.org/abs/1203.4236}{{\ttfamily arXiv:1203.4236 [hep-th]}}.

\bibitem{Donos:2012yi}
A.~Donos, J.~P. Gauntlett, and C.~Pantelidou, ``{Semi-local quantum criticality
  in string/M-theory},'' \href{http://dx.doi.org/10.1007/JHEP03(2013)103}{{\em
  JHEP} {\bfseries 1303} (2013) 103},
\href{http://arxiv.org/abs/1212.1462}{{\ttfamily arXiv:1212.1462 [hep-th]}}.

\bibitem{Iqbal:2008by}
N.~Iqbal and H.~Liu, ``{Universality of the hydrodynamic limit in AdS/CFT and
  the membrane paradigm},''
  \href{http://dx.doi.org/10.1103/PhysRevD.79.025023}{{\em Phys.Rev.}
  {\bfseries D79} (2009) 025023},
\href{http://arxiv.org/abs/0809.3808}{{\ttfamily arXiv:0809.3808 [hep-th]}}.

\bibitem{Blake:2013owa}
M.~Blake, D.~Tong, and D.~Vegh, ``{Holographic Lattices Give the Graviton a
  Mass},''
\href{http://arxiv.org/abs/1310.3832}{{\ttfamily arXiv:1310.3832 [hep-th]}}.

\bibitem{Goldstein:2009cv}
K.~Goldstein, S.~Kachru, S.~Prakash, and S.~P. Trivedi, ``{Holography of
  Charged Dilaton Black Holes},''
  \href{http://dx.doi.org/10.1007/JHEP08(2010)078}{{\em JHEP} {\bfseries 1008}
  (2010) 078},
\href{http://arxiv.org/abs/0911.3586}{{\ttfamily arXiv:0911.3586 [hep-th]}}.

\bibitem{Cadoni:2009xm}
M.~Cadoni, G.~D'Appollonio, and P.~Pani, ``{Phase transitions between
  Reissner-Nordstrom and dilatonic black holes in 4D AdS spacetime},''
  \href{http://dx.doi.org/10.1007/JHEP03(2010)100}{{\em JHEP} {\bfseries 1003}
  (2010) 100},
\href{http://arxiv.org/abs/0912.3520}{{\ttfamily arXiv:0912.3520 [hep-th]}}.

\bibitem{Charmousis:2010zz}
C.~Charmousis, B.~Gouteraux, B.~Kim, E.~Kiritsis, and R.~Meyer, ``{Effective
  Holographic Theories for low-temperature condensed matter systems},''
  \href{http://dx.doi.org/10.1007/JHEP11(2010)151}{{\em JHEP} {\bfseries 1011}
  (2010) 151},
\href{http://arxiv.org/abs/1005.4690}{{\ttfamily arXiv:1005.4690 [hep-th]}}.

\bibitem{Mateos:2011tv}
D.~Mateos and D.~Trancanelli, ``{Thermodynamics and Instabilities of a Strongly
  Coupled Anisotropic Plasma},''
  \href{http://dx.doi.org/10.1007/JHEP07(2011)054}{{\em JHEP} {\bfseries 1107}
  (2011) 054},
\href{http://arxiv.org/abs/1106.1637}{{\ttfamily arXiv:1106.1637 [hep-th]}}.

\bibitem{Iizuka:2012wt}
N.~Iizuka and K.~Maeda, ``{Study of Anisotropic Black Branes in Asymptotically
  anti-de Sitter},'' \href{http://dx.doi.org/10.1007/JHEP07(2012)129}{{\em
  JHEP} {\bfseries 1207} (2012) 129},
\href{http://arxiv.org/abs/1204.3008}{{\ttfamily arXiv:1204.3008 [hep-th]}}.

\bibitem{Faulkner:2009wj}
T.~Faulkner, H.~Liu, J.~McGreevy, and D.~Vegh, ``{Emergent quantum criticality,
  Fermi surfaces, and AdS(2)},''
  \href{http://dx.doi.org/10.1103/PhysRevD.83.125002}{{\em Phys.Rev.}
  {\bfseries D83} (2011) 125002},
  \href{http://arxiv.org/abs/0907.2694}{{\ttfamily arXiv:0907.2694 [hep-th]}}.

\bibitem{Huijse:2011ef}
L.~Huijse, S.~Sachdev, and B.~Swingle, ``{Hidden Fermi surfaces in compressible
  states of gauge-gravity duality},''
  \href{http://dx.doi.org/10.1103/PhysRevB.85.035121}{{\em Phys.Rev.}
  {\bfseries B85} (2012) 035121},
\href{http://arxiv.org/abs/1112.0573}{{\ttfamily arXiv:1112.0573
  [cond-mat.str-el]}}.

\bibitem{Ogawa:2011bz}
N.~Ogawa, T.~Takayanagi, and T.~Ugajin, ``{Holographic Fermi Surfaces and
  Entanglement Entropy},''
  \href{http://dx.doi.org/10.1007/JHEP01(2012)125}{{\em JHEP} {\bfseries 1201}
  (2012) 125},
\href{http://arxiv.org/abs/1111.1023}{{\ttfamily arXiv:1111.1023 [hep-th]}}.

\bibitem{Donos:2012yu}
A.~Donos, J.~P. Gauntlett, J.~Sonner, and B.~Withers, ``{Competing orders in
  M-theory: superfluids, stripes and metamagnetism},''
  \href{http://dx.doi.org/10.1007/JHEP03(2013)108}{{\em JHEP} {\bfseries 1303}
  (2013) 108},
\href{http://arxiv.org/abs/1212.0871}{{\ttfamily arXiv:1212.0871 [hep-th]}}.

\bibitem{2003Natur.425..271M}
D.~v.~d. {Marel}, H.~J.~A. {Molegraaf}, J.~{Zaanen}, Z.~{Nussinov},
  F.~{Carbone}, A.~{Damascelli}, H.~{Eisaki}, M.~{Greven}, P.~H. {Kes}, and
  M.~{Li}, ``{Quantum critical behaviour in a high-T$_{c}$ superconductor},''
  \href{http://dx.doi.org/10.1038/nature01978}{{\em Nature} {\bfseries 425}
  (Sept., 2003) 271--274},
  \href{http://arxiv.org/abs/arXiv:cond-mat/0309172}{{\ttfamily
  arXiv:cond-mat/0309172}}.

\bibitem{Gouteraux:2014hca}
B.~Gouteraux, ``{Charge transport in holography with momentum dissipation},''
\href{http://arxiv.org/abs/1401.5436}{{\ttfamily arXiv:1401.5436 [hep-th]}}.

\bibitem{Gubser:2009qt}
S.~S. Gubser and F.~D. Rocha, ``{Peculiar properties of a charged dilatonic
  black hole in $AdS_5$},''
  \href{http://dx.doi.org/10.1103/PhysRevD.81.046001}{{\em Phys.Rev.}
  {\bfseries D81} (2010) 046001},
\href{http://arxiv.org/abs/0911.2898}{{\ttfamily arXiv:0911.2898 [hep-th]}}.

\end{thebibliography}\endgroup
\end{document}